\documentclass[amssymb,amsmath,preprint]{revtex4}
\usepackage{graphicx}
\usepackage{hyperref}
\makeatletter
\def\@cite#1#2{{\m@th\upshape\bfseries%
[{#1\if@tempswa{\m@th\upshape\mdseries, #2}\fi}]}}
\makeatother
\newtheorem{thm}{Theorem}[section]
\newtheorem{lem}[thm]{Lemma}
\newtheorem{cor}[thm]{Corollary}
\newtheorem{prop}[thm]{Proposition}
\newtheorem{rem}[thm]{Remark}

\newtheorem{defn}[thm]{Definition}

\newtheorem{eg}[thm]{Example}

\newtheorem{prob}[thm]{Problem}
\newcommand{\Prf}{\noindent\textbf{Proof.\ }}
\newcommand{\bx}{\strut\hfill$\blacksquare$\medbreak}

\newcommand{\ca}{\mathrm{C}^*}

\newenvironment{sbmatrix}{\left[\begin{smallmatrix}}{\end{smallmatrix}\right]}
\newenvironment{spmatrix}{\left(\begin{smallmatrix}}{\end{smallmatrix}\right)}
\newcommand{\td}{\widetilde}

\newcommand{\bbC}{{\mathbb{C}}}

 \newcommand{\A}{{\mathcal{A}}}
 \newcommand{\B}{{\mathcal{B}}}
 \newcommand{\C}{{\mathcal{C}}}
 \newcommand{\D}{{\mathcal{D}}}

\renewcommand{\H}{{\mathcal{H}}}

 \newcommand{\M}{{\mathcal{M}}}
 
\renewcommand{\O}{{\mathcal{O}}}

\renewcommand{\S}{{\mathcal{S}}}

\newcommand{\upchi}{{\raise.35ex\hbox{$\chi$}}}
\newcommand{\qand}{\quad\text{and}\quad}

\newcommand{\qfor}{\quad\text{for}\quad}
\newcommand{\qforsome}{\quad\text{for some}\quad}

\newcommand{\qforal}{\quad\text{for all}\quad}

\newcommand{\Alg}{\operatorname{Alg}}

\newcommand{\rank}{\operatorname{rank}}

\newcommand{\spn}{\operatorname{span}}

\newcommand{\fix}{\operatorname{Fix}}
\def\bra#1{\langle #1|}
\def\ket#1{|#1 \rangle}
\def\one{{\mathchoice{\rm 1\mskip-4mu l}{\rm 1\mskip-4mu l}{\rm 1\mskip-4.5mu l}{\rm
1\mskip-5mu l}}}

\newcommand{\fixed}{\fix (\Phi)}
\newcommand{\sumkn}{\sum_{k=1}^n}

\newcommand{\bofh}{\B(\H)}

\begin{document}

\title{Noiseless subsystems and the structure of the
commutant in quantum error correction}
%
\author{John A. Holbrook$^1$, David W. Kribs$^{1,3}$ and Raymond Laflamme$^{2,3}$}

\affiliation{$^1$Department of Mathematics and Statistics, University of
Guelph, Guelph, Ontario, Canada  N1G 2W1. \\ $^2$Institute for Quantum
Computing,
University of Waterloo, Waterloo, ON, CANADA N2L 3G1. \\ $^3$Perimeter Institute
for Theoretical Physics, 35 King St. North, Waterloo, ON, CANADA N2J 2W9.}
\email{jholbroo@uoguelph.ca; dkribs@uoguelph.ca;  laflamme@iqc.ca}

%
\begin{abstract}
The effect of noise on a quantum system can be described by
a set of operators obtained from the interaction Hamiltonian.
Recently it has been shown that generalized quantum error correcting
codes can be derived by studying the algebra of this set of
operators. This led to the discovery of noiseless subsystems.
They are described by a set of operators obtained from the commutant of
the noise generators.  In this paper we derive a general method to
compute the structure of this commutant in the case of unital noise.
\end{abstract}
\maketitle

\section{Introduction}\label{S:intro}

Quantum mechanics promises to manipulate information for
communication, cryptography and computation in a way fundamentally
different from its classical counterpart\cite{NC}. Although it is
possible to manipulate small quantum systems in the laboratory,
the task to do so for large ones is daunting, especially because
in absence of control of noise and imperfection of realistic
devices the quantum properties of the state are destroyed. Quantum
error correction methods have recently been discovered which
protect quantum information against corruption. In particular it
was shown that if the error rate is below a certain threshold
\cite{AB,Kit,KLZ,Preskillplenum} then it is possible implement
quantum error correction and obtain a fixed probability of success
with only a polynomial amount of resources. The work on the
accuracy threshold assumes certain error models such as
independent or quasi independent errors. The theorem shows that
under reasonable assumptions, imperfections and imprecisions of
realistic devices do not create fundamental objections to scalable
quantum information processing.

The estimated value of the error
threshold is  small and, today, an extraordinary challenge
for experimentalists. From the experimental point of view there is a need to
better understand the details of error models of quantum devices, the
form of the  Kraus noise operators which occur during the evolution of a
quantum computation or during the transmission in a quantum channel, the
correlation between
errors, their strength, etc. From the theoretical side we need to find
efficient ways to estimate the error model, determine the threshold
for various physically relevant error models and find ways to achieve
better  threshold. This can be done by optimizing quantum error correction
procedures or simply finding better ways to protect quantum information.

Recently a unified view of error correction was given which
brought together quantum error correcting codes and noiseless
(decoherence-free) subspaces \cite{DG,KLV,LCW,Zan,ZR}.  This
investigation led to the discovery of noiseless subsystems.  A
quantum bit, called a qubit,  is described by a unit vector in a
2-dimensional Hilbert space. Often this 2-dimensional space is
directly associated with a two-level physical system such as the
ground and excited states of an atom, a nuclear spin pointing up
or down, etc.  Noiseless subsystems are instead described by a set
of operators such as the Pauli matrices, a more general concept
than a subspace of a Hilbert space\cite{KLV}. An example where the
difference can be appreciated is in cases where the systems at
hand are subject to noise with some symmetries and we are
interested in a fraction of the operators acting on the whole
Hilbert space of the systems.  For example, the polarization of
photons moving in an optic fiber undergo collective rotation. When
a symmetry of the noise exists, it is possible to find conserved
quantities and thus invariant, or fixed, operators. These
operators form the noise commutant $\A^\prime$ of the noise.

When the channel determined by the transmission is unital (i.e.  the unit
matrix is preserved), the commutant is a $\ca$-algebra \cite{Kchannel}.
Such an algebra has a block diagonal matrix form
\begin{eqnarray}\label{decompform}
\A^\prime = \begin{sbmatrix}
\begin{array}{|c|}
\hline
\ast \\
\hline
\end{array}
  &   &  &0  \\
 &
\begin{array}{|c|}
\hline
\ast \\
\hline
\end{array}
     &  &  \\
 & & \ddots & \\
 0& & &
\begin{array}{|c|}
\hline
\ast \\
\hline
\end{array}
\end{sbmatrix},
\end{eqnarray}
where the compression to each block forms a full matrix algebra.
But some of these blocks may be `linked' in a sense that we make
precise below. Once the block matrix form and the various links
between blocks have been found, the explicit form of the algebra
is revealed. In particular, there are unique positive integers
$n_k, m_k \geq 1$ for $k=1,\ldots,d$ such that the algebra is
unitarily (spatially) equivalent to an  orthogonal direct sum
\[
\A^\prime \simeq \big(\M_{n_1} \otimes \one_{m_1}\big) \oplus \ldots \oplus
 \big(\M_{n_d} \otimes \one_{m_d}\big),
\]
where $\M_{n_k}$ is the full $n_k \times n_k$ matrix algebra on
$n_k$-dimensional Hilbert space and $\one_{m_k}$ is the identity
operator on $m_k$-dimensional space. A tensor product $\M_{n_k}
\otimes \one_{m_k}$ in this decomposition corresponds to $m_k$
blocks of $n_k \times n_k$ matrices in (\ref{decompform}) which
are linked, and is referred to as an ampliation of $\M_{n_k}$ when
$m_k \geq 2$. From the representation theory perspective, the
integers $m_k$ correspond to the multiplicities of the identity
representations of $\M_{n_k}$ which determine the structure of
$\A^\prime$.

If there are invariant operators which can be decomposed into projectors of rank
one, so they are supported on blocks with $n_k =1$, then  they correspond to invariant subspaces
and the usual notion of
qubits. Otherwise the invariant operators correspond to noiseless
subsystems on higher dimensional spaces. This can be read off the structure of the commutant.
For instance, in the above discussion there will be a copy of every operator from $\M_{n_k}$
contained in the $\M_{n_k} \otimes \one_{m_k}$ component of the decomposition.
Thus the structure of the commutant indicates how quantum information can
be encoded to be preserved under the influence of the noise.

In this paper we study how to find the structure of the commutant
once the noise (Kraus) operators are known.  A constructive proof
of this structure is given in the case of unital evolution.  The
paper is structured as follows. In sections 2 and 3 we review
background concepts. In particular, we recall basic properties of
quantum channels and results from \cite{Kchannel} on the fixed
point set of a unital channel. Section~4 contains the core
theoretical component of the paper. Based on operator theory and
operator algebra methods, we present a detailed  algorithm for
computing the noise commutant of a unital channel. We apply this
method in Section~5 to find the commutant for a number of simple
illustrative examples which are related to the phase damping
channel \cite{FVHTC,KBLW,NC}. Section~6 contains a short
discussion on how to turn a non-unital channel into a unital one,
when the noise operators satisfy certain constraints. In the final
three sections we conduct an  analysis of special cases of
collective noise channels which arise from collective rotation
\cite{Fi,KLV,KBLW,Lnoiseless,VFPKLC,VKL,Zan}. Specifically,
section~7 includes an introductory discussion on this class, and
in the final two sections we use the algorithm to explicitly
compute the commutant for the 3-qubit
\cite{KBLW,Lnoiseless,VFPKLC,VKL} and 4-qubit channels from this
class.

\section{Channel Preliminaries}\label{S:channelprelim}

Let $\H$ be a finite dimensional Hilbert space and let $\bofh$ be the set of all bounded
(continuous) operators acting on $\H$. A {\it quantum channel} (or {\it quantum evolution}) on
$\H$
is a linear map
$\Phi : \bofh \rightarrow \bofh$ which is completely positive and trace preserving. We mention
the texts \cite{Paulsentext,Paulsentext2} for basic properties of completely positive maps, and
\cite{NC} for
an introduction to channels in quantum information theory.

It is well-known \cite{Choi,Kraus} that to every completely positive map $\Phi$ there
corresponds a     (non-unique) family of operators $\{ A_1, \ldots, A_n\}$ in $\bofh$ which
determine the map $\Phi$ through the equation
\begin{eqnarray}\label{standardform}
\Phi(T) = \sumkn A_k T A_k^\dagger \qfor T\in\bofh.
\end{eqnarray}
Trace preservation of $\Phi$ is equivalent to these operators
satisfying
\begin{eqnarray*}
 \sumkn A_k^\dagger A_k = \one,
\end{eqnarray*}
where $\one$ is the identity operator on $\H$.
(This is also equivalent to the `dual map' of $\Phi$ being unital.)
We say $\Phi$ is {\it unital} if also,
\[
\Phi(\one) = \sumkn A_k A_k^\dagger = \one.
\]
For a quantum channel $\Phi$, the family of operators $\{A_k \}$ determining $\Phi$ as in
(\ref{standardform}) are called the {\it Kraus noise operators} for the channel.

The fixed point set for $\Phi$ will be denoted by
\[
\fixed = \big\{ T\in \bofh \, : \, \Phi (T) = T \big\}.
\]
We shall also let $\A$ be the so-called {\it interaction algebra}
\cite{KLV} generated by the noise operators,
\begin{eqnarray*}
\A = \Alg \{A_1, \ldots, A_n \}.
\end{eqnarray*}
In other words, $\A$ is the set of all polynomials  in the
typically noncommuting variables $A_1, \ldots, A_n$. Note that as
an application of the Cayley-Hamilton theorem from linear algebra,
the degree of any such polynomial may be reduced below some
uniform bound. The {\it noise commutant} $\A^\prime$ of $\A$ is
the set of all operators which commute with every operator in
$\A$; it is the algebra
\begin{eqnarray*}
\A^\prime &=& \big\{ T\in\bofh \,\,\, : \,\,\, TA=AT \qforal A\in\A \big\}
\\
&=& \big\{ T\in\bofh \,\,\, : \,\,\, TA_k =A_k T \qfor k=1,\ldots, n
\big\}
\end{eqnarray*}

\begin{rem}
{\rm The interaction algebra, and hence the noise commutant, are
usually defined as $\dagger$-algebras \cite{KLV}. However, in the
case considered here, that of unital noise, the discussion below
shows that our `nonselfadjoint' definitions are actually a red
herring, the algebras turn out to be selfadjoint.}
\end{rem}

In this paper, we shall focus our attention on {\it unital quantum
channels}. These are sometimes called {\it bistochastic channels} in the literature.
For such channels, it was proved in \cite{Kchannel} that:
\begin{itemize}
\item[$(i)$] $\A$ is a $\dagger$-algebra; in other words, $A$
belongs to $\A$ if and only if $A^\dagger$ belongs to $\A$.
Furthermore, the algebra $\A$ is independent of the choice of
noise operators which determine the channel as in
(\ref{standardform}); the same algebra is obtained whatever the
choice of operators $\{ A_k \}$. \item[$(ii)$] The commutant
$\A^\prime$ is a $\dagger$-algebra which is equal to the fixed
point set of $\Phi$,
\[
\A^\prime = \fixed.
\]
\end{itemize}

These results are recalled in the next section. Taken together,
they may be regarded as a structure theorem for the fixed point
set of a unital quantum channel. In particular, the fixed point
set $\fixed$ (which is just a $\dagger$-closed subspace for
general completely positive maps) is itself an algebra, hence
closed under multiplication, equal to the noise commutant. But
there is a tight decomposition theory for finite dimensional
$\dagger$-algebras (see section~\ref{S:algorithm} for details).
This motivates the following fundamental problem which we consider
in this paper:
\begin{prob}
Given a unital quantum channel $\Phi$, explicitly determine the
algebra structure of the noise commutant $\fixed = \A^\prime$.
\end{prob}
There are compelling  reasons in quantum information theory for
wishing to solve this problem. For instance, in the noiseless
subsystem method of quantum error correction
\cite{DG,KLV,Lnoiseless,LCW,VFPKLC,ZR}, the structure of the noise
commutant can be used to encode quantum information which is
immune to the noise of the corresponding channel.

\section{Structure Theorem}\label{S:structure}

In \cite{Kchannel} it was discovered that many  properties of a channel are married to properties
of the noise operators.
Recall that an orthogonal
projection $P$ on $\H$ {\it reduces} an operator $T$ in $\bofh$ when both subspaces $P \H$
and $(\one -P)\H = P^\perp \H$ are invariant for $T$; that is, $T(P\H)
\subseteq P\H$ and
$T(P^\perp\H) \subseteq P^\perp\H$. This is equivalent to the identity $TP = PT$ being
satisfied, or $T$ having a block diagonal matrix form
$T = \begin{sbmatrix}
B & 0 \\
0 & C
\end{sbmatrix}$
with respect to the orthogonal spatial decomposition $\H = P \H \oplus P^\perp \H$.
Further recall the operator inequality $P\leq Q$ for a pair of self-adjoint operators on $\H$
means that $\bra{P\xi}\ket{\xi} \leq \bra{Q\xi}\ket{\xi}$ for all $\xi\in\H$.
The key point in the following result is that invariant subspaces for the noise operators of a
unital channel are always reducing subspaces.

\begin{lem}\label{structurelemma}
\cite{Kchannel} Let $\Phi : \bofh \rightarrow \bofh$ be a unital completely positive map and let
$P$ be a projection on $\H$. Consider the following conditions.
\begin{itemize}
\item[$(i)$] $\Phi (P) = P$.
\item[$(ii)$] The subspace $P \H$ reduces $A_1, \ldots, A_n$;
\[
A_k P = P A_k \qfor 1 \leq k \leq n.
\]
\item[$(iii)$] $\Phi (P) \leq P$.
\item[$(iv)$] The subspace $P \H$ is invariant for  $A_1, \ldots, A_n$;
\[
A_k P = P A_k P \qfor 1 \leq k \leq n.
\]
\item[$(v)$] $\Phi (P) \geq P$.
\item[$(vi)$] The subspace $P \H$ is invariant for  $A_1^\dagger, \ldots, A_n^\dagger$;
\[
P A_k  = P A_k P \qfor 1 \leq k \leq n.
\]
\end{itemize}
Then the pairs of conditions $((i),(ii))$,    $((iii),(iv))$, and
$((v),(vi))$ are each  equivalent. Moreover, if  $\Phi$ is a unital channel then all six conditions
are equivalent to each other.
Further,  given a projection $P$ and a unital channel $\Phi$,  conditions $(ii)$, $(iv)$ and $(vi)$
hold for one choice  of noise operators for  $\Phi$ if and only if they hold for all choices.
\end{lem}

Let us outline the proof of $(iv) \Rightarrow (ii)$ for a unital channel $\Phi$, since this is the
aspect of
Lemma~\ref{structurelemma} used in the proof of Theorem~\ref{structure}.
Suppose $(iv)$ holds for a projection $P$ in $\bofh$. Then
\[
0\leq \Phi(P) = \sum_{i}A_i P A_i^\dagger = P \Phi(P) P \leq P \Phi(\one)P = P\one P = P.
\]
As $0 \leq \Phi(P) \leq P$, trace preservation can be seen to imply through an operator theory
argument that $\Phi(P) = P$. More generally, if $R$ is a positive contraction operator with $R
\leq P$ for some projection $P$ and $R$ has trace equal to the trace of $P$, then in fact $R =P$.
Thus
$\Phi(P^\perp) = \Phi (\one - P) = P^\perp$ since $\Phi$ is unital, and for each $j$
\begin{eqnarray*}
0 &\leq& \big(  P^\perp A_j^\dagger P \big)^\dagger \big( P^\perp A_j^\dagger P \big) \\
&\leq &
\sum_i \big(  P^\perp A_i^\dagger P \big)^\dagger \big( P^\perp A_i^\dagger P \big)
= P \Phi (P^\perp) P =  P P^\perp P = 0.
\end{eqnarray*}
Thus, $P^\perp A_j^\dagger P = 0$ for all $j$, and hence $P\H$ is a reducing subspace
for $A_1, \ldots, A_n$.

We mention that this lemma gives a simple characterization of rank one projections
fixed by a unital channel.

\begin{cor}\label{rankone}
Let $P = \ket{\xi}\bra{\xi}$ be a rank one projection in $\bofh$.
Then the following are equivalent:
\begin{itemize}
\item[$(i)$] $\Phi(P) = P$.
\item[$(ii)$] The range vector $\xi$ for $P$ is a joint eigenvector for the noise operators $A_1,
\ldots, A_n$;
\[
A_k \,\xi = \lambda_k \, \xi \qforsome \lambda_k \in \bbC \qand
k=1,\ldots, n. \]
\end{itemize}
\end{cor}

\Prf
Condition $(ii)$ here is simply a restatement of $(iv)$ from the lemma, in the special case of a
rank
one projection.
\bx

We recall some terminology from linear algebra  before proceeding.

\begin{defn}
{\rm
Let  $T$ be a normal operator in $\bofh$, in other words $TT^\dagger = T^\dagger T$. Then  a
{\it spectral projection} for $T$ is a
projection
onto an eigenspace for $T$; that is, it is a projection onto a
subspace of the form $\{ \xi \in \H \, : \, T\xi = \lambda \xi \}$ for
some $\lambda\in \bbC$.
The Spectral Theorem from linear algebra states that every normal operator $T$ decomposes as
a sum $T = \lambda_1 P_1 + \ldots + \lambda_r P_r$, where $\lambda_i \in \bbC$ are the
distinct eigenvalues for $T$ and $P_i$ the corresponding spectral projections (which necessarily
have mutually orthogonal ranges).
}
\end{defn}

Clearly the fixed point set $\fixed$ contains the noise commutant
$\A^\prime$;  if $T$ belongs to $\A^\prime$ then
\[
\Phi (T) = \sumkn A_k T A_k^\dagger = T \Big( \sumkn A_k A_k^\dagger \Big) = T \Phi(\one)
= T.
\]
The converse inclusion holds for unital channels. We mention that
the following theorem was partly motivated by work from \cite{DKS}
on infinite dimensional operator algebras and dilation theory. The
proof in \cite{Kchannel} is somewhat technical, but after
submission we discovered a simpler proof of this fact, and hence,
for completeness, we shall present it here.

\begin{thm}\label{structure}
\cite{Kchannel} Let $\Phi : \bofh \rightarrow \bofh$ be a unital
quantum channel determined as in $(\ref{standardform})$ by
operators $\{ A_1, \ldots, A_n \}$. Then the fixed point set for
$\Phi$ is a $\dagger$-algebra and is equal to the noise commutant,
\[
\fixed = \A^\prime = \big\{ A_1, \ldots , A_n \big\}^\prime,
\]
Further, the interaction algebra $\A$ generated by $A_1, \ldots,
A_n$ is a $\dagger$-algebra which depends only on $\Phi$ via the
equation
\[
\A = \fixed^\prime.
\]
\end{thm}

\Prf
Notice that the theorem will be proved if it is shown that $\A^\prime = \fixed$. Indeed,
$\A^\prime$ is an algebra by definition and $\fixed$ is $\dagger$-closed since $\Phi$ is a
positive map.
Moreover, if this identity holds, the algebra $\A$ depends only on $\Phi$, not the choice of noise
operators,
since $\A = (\A^\prime)^\prime = \fixed^\prime$. (This is a special case of von Neumann's
double commutant theorem from operator algebras.)

As observed above, the commutant $\A^\prime$ is
contained in $\fix(\Phi)$ because $\Phi$ is unital. To see the converse,
first observe that if the only fixed points are scalars, then $\bbC \one
\subseteq \A^\prime \subseteq \fixed = \bbC \one$, and the two sets
coincide. Otherwise,  let $T$ be a
non-scalar operator in $\fixed$.
Without loss of generality we
may assume that $T$ is self-adjoint; in other words, $T  = T^\dagger $. Indeed,  $\Phi
(T^\dagger) = \Phi(T)^\dagger =T^\dagger$ by positivity of $\Phi$, and hence both the real,
${\rm Re}\, (T)= \frac{1}{2} (T+T^\dagger) $, and imaginary, ${\rm Im} \, (T) =
\frac{1}{2i}(T - T^\dagger)$, parts of $T = {\rm Re}\, (T) + i \,{\rm Im} \, (T)$ are fixed by
$\Phi$. Since $T$ is non-scalar, at least one of these self-adjoint operators is non-scalar.
Furthermore, by  replacing $T$ with $T + ||T||
\one$, we may assume $T \geq 0$ is a positive operator. Let
$0 \leq \lambda_1 < \lambda_2 < \ldots < \lambda_r$ be the eigenvalues for
$T$, and $P_1,\ldots ,P_r$ the corresponding spectral projections. Then $T
= \lambda_1 P_1 + \ldots + \lambda_r P_r$ by the Spectral Theorem, and
$0 \leq T \leq ||T||\one = \lambda_r \one$. Let $\H_r = P_r \H = \{ \xi\in
\H : T\xi = \lambda_r \xi\}$ be the eigenspace for the extremal
eigenvalue $\lambda_r$. If $\xi$ is a non-zero vector in $\H_r$, observe
that
\begin{eqnarray*}
 \lambda_r \bra{\xi}\ket{\xi} = \bra{T \xi}\ket{\xi} & = & \sum_{k=1}^n \bra{A_k T
A_k^\dagger \xi}\ket{\xi} \\
    & = & \sum_{k=1}^n \bra{T A_k^\dagger \xi}\ket{A_k^\dagger \xi}         \\
    & \leq& \lambda_r \sum_{k=1}^n \bra{A_k^\dagger \xi}\ket{A_k^\dagger \xi} =  \lambda_r
\bra{\xi}\ket{\xi}.
\end{eqnarray*}
The only way this can happen is if each of the inequalities is actually an
equality; in particular, $A_k^\dagger \xi$ belongs to $\H_r$ for $k= 1,
\ldots, n$. Thus $\H_r = P_r \H$ is an invariant subspace for
$\A^\dagger$, and so it is a reducing subspace for $\A$ by
Lemma~\ref{structurelemma}. Therefore, it follows that $P_r$ belongs to
the commutant $\A^\prime$, and hence to the fixed point set $\fixed$.
Thus the self-adjoint operator $T - \lambda_r P_r$ belongs to $\fixed$,
and we may iterate this argument to find that each of $P_1, \ldots ,P_r$
belongs to $\A^\prime$. Hence $T$ belongs to $\A^\prime$, as required.

We have shown that every $T = T^\dagger$ in $\fixed$ also belongs to $\A^\prime$.
But $\fixed$ is a self-adjoint subspace, and hence spanned by its
self-adjoint part. It follows that the fixed point set $\fixed$ is contained in the commutant
$\A^\prime$, and this completes the proof.
\bx

\section{Computing the Commutant}\label{S:algorithm}

In this section, using operator theory and operator algebra
techniques, we present a constructive process for explicitly
computing the noise commutant $\A^\prime$ for a unital quantum
channel.  This will be accomplished by first identifying the
$\dagger$-algebra structure of the interaction algebra $\A$. We
begin with a short discussion on the decomposition theory for
$\dagger$-algebras.

A fundamental result from the theory of operator algebras is that every $\dagger$-algebra $\A$ is
$\dagger$-isomorphic (an isomorphism which preserves adjoints) to a unique orthogonal direct
sum of full matrix algebras. Specifically, there are positive integers $n_k \geq 1$ such that $\A$
is $\dagger$-isomorphic to
\begin{eqnarray}\label{daggerform}
\A \simeq \M_{n_1} \oplus \ldots \oplus \M_{n_d}.
\end{eqnarray}
We mention the texts \cite{Arvinvite,byeg,Tak} for basics of $\dagger$-algebras, or
equivalently, finite dimensional $\ca$-algebras.
Another important result is that every representation of such an algebra decomposes into a direct
sum of multiples of the identity representation on each summand. As a consequence, up to
unitary equivalence $\A$ is given by a unique direct sum of `ampliated' full matrix algebras,
\begin{eqnarray}\label{spatialform}
\A \simeq  (\M_{n_1} \otimes \one_{m_1}) \oplus \ldots \oplus (\M_{n_d} \otimes
\one_{m_d}).
\end{eqnarray}
In other words, there is a unitary operator $U$ such that
$U^\dagger \A U$ is equal to the decomposition in
(\ref{spatialform}). See also \cite{VKL,ZL,Zan,KBLW,Lnoiseless}
for discussions of these decompositions in connection with quantum
information processing.

While it is important to know the $\dagger$-isomorphism class of
$\A$ as in (\ref{daggerform}), we really need the structure of
$\A$ spatially as in (\ref{spatialform}). More to the point, we
need to identify the precise spatial locations of the matrix
blocks $\M_{n_k} \otimes \one_{m_k}$. The integers $m_k \geq 1$
are the multiplicities of the identity representation appearing in
each summand associated with the corresponding representation of
the $\dagger$-isomorphism class model in (\ref{daggerform}). The
invariants $n_k, m_k$ are part of what is known as the Wedderburn
structure theory for such algebras. When some $m_k \geq 2$, we say
$\M_{n_k} \otimes \one_{m_k}$ is an {\it ampliation} of
$\M_{n_k}$. The projections in $\A$ which, under the unitary
equivalence (\ref{spatialform}), correspond to the identity
operators on the blocks $\M_{n_k} \otimes \one_{m_k}$ in the
decomposition (\ref{spatialform}) are called the {\it minimal
central projections} for $\A$. Each of these projections will be a
sum of projections onto some, but perhaps not all, of the matrix
blocks of the same size in the block diagonal form
(\ref{decompform}).

The projections onto the individual matrix blocks in
(\ref{decompform}) will be minimal inside $\A^\prime$, in other
words minimal $\A$-reducing, but may not belong to the algebra
$\A$. This is precisely the subtlety we must deal with in this
process; an identification of `links' between these minimal
projections. Thus, broadly speaking, we must first identify the
maximal (here meaning that the projections sum to the identity
operator) family of non-zero minimal  $\A$-reducing projections
with mutually orthogonal ranges (part~I below); it is easy to see
that such a family of projections is unique. Then we must detect
any links between these projections (part~II below) in the
following sense.

\begin{defn}
{\rm
Let $\{P_j\}$ be the (unique) maximal family of non-zero mutually orthogonal minimal
$\A$-reducing  projections.  A subset $\{Q_k\}_{k\in\S}$ of this family is {\it linked} in $\A$ if
the following two conditions hold.
\begin{itemize}
\item[$(i)$] The projection $Q = \sum_{k\in\S} Q_k$ belongs to $\A$.
\item[$(ii)$] If
$\S_1$ is a proper subset of $\S$, then the projection $\sum_{k\in\S_1} Q_k$ does not belong to
$\A$.
\end{itemize}
}
\end{defn}

Let $\{ P_j : j\in \S_k \}_k$ be the linked subsets of projections
from $\{P_j\}$. It follows that each algebra $\A \big(
\sum_{j\in\S_k} \oplus P_j \big)$ forms a {\it subalgebra} of
$\A$. Notice that, in terms of (\ref{spatialform}), the
cardinality of the sets $\S_k$ are equal to the ampliation
multiplicities $m_k$.

Once the $\dagger$-algebra structure of $\A$ has been identified
up to unitary equivalence, we can easily deduce the
$\dagger$-algebra structure of $\A^\prime = \fixed$. For instance,
if $\Phi$ is a channel and the spatial structure of $\A$ is given
by (\ref{spatialform}), then
\[
\fixed = \A^\prime \simeq \sum_k \oplus (\M_{n_k} \otimes
\one_{m_k})^\prime \simeq \sum_k \oplus (\one_{n_k} \otimes
\M_{m_k}).
\]
Furthermore, the algorithm presented below will allow us to
explicitly identify the locations of the matrix blocks, possibly
ampliated, which belong to the noise commutant.

We shall work our way through the finer points of the process,
then finish by summarizing the main points. For the rest of this
section we let $\Phi$ be a given unital quantum channel acting on
$\bofh$, with noise operators $\{A_1, \ldots, A_n\}$, and
interaction algebra  $\A = \Alg \{A_1, \ldots, A_n\}$. The
following assumption gives us the starting point.
\[
\begin{array}{cl}
\mbox{{\it Assumption:}} & \mbox{{\it The noise commutant
$\A^\prime = \fix(\Phi)$ has been}}  \\
 & \mbox{{\it computed as a linear span.}}
\end{array}
\]
The algorithm will take us from knowing the vector space structure of $\A^\prime$, to
identifying the algebra structure of $\A$, and hence of
$\A^\prime$. There are two broad
components to the process:
\begin{itemize}
\item[] {\bf Part I.} Find  the family of non-zero projections
inside the noise commutant $\A^\prime$ which have mutually
orthogonal ranges, sum to the identity operator, and are each
minimal inside $\A^\prime$. \item[]{\bf Part II.} Given the family
of projections $\{ P_j \}$ from Part I, find the minimal central
projections for $\A$ by computing all links between the $P_j$.
This will give the spatial $\dagger$-algebra structure of $\A$,
and hence of  $\fixed = \A^\prime$.
\end{itemize}

\subsection{Part I}\label{sS:partI}

As $\Phi$ is unital, $\A^\prime = \fixed $ always  contains the scalar operators. If
$\A^\prime =
\bbC \one$, then the singleton $\{ \one \}$ is the only (non-zero) minimal
central projection for $\A$. Otherwise,
let $T = T^\dagger$ belong to $\A^\prime \setminus \bbC \one$.
The spectral projections for $T$ provide the first step toward the maximal family.

\begin{lem}\label{spectralprojns}
The following are equivalent for an operator $T =T^\dagger $ in $\bofh$:
\begin{itemize}
\item[$(i)$] $\Phi(T) = T$.
\item[$(ii)$] Every spectral projection $P$ of $T$ satisfies $\Phi (P) = P$.
\item[$(iii)$]  Every spectral projection $P$ of $T$ belongs to $\A^\prime$.
\end{itemize}
\end{lem}

\Prf
Condition $(ii)$ and $(iii)$ are equivalent by Theorem~\ref{structure}.
The equivalence of the first two conditions immediately follows from
standard operator algebra  for
$\dagger$-algebras. We provide a proof for the sake of brevity.
The implication $(ii)\Rightarrow (i)$ follows from the Spectral Theorem of
linear algebra;
$T$ decomposes as a sum $T = \lambda_1 P_1 + \ldots + \lambda_r P_r$ where $\lambda_1,
\ldots, \lambda_r$ are the distinct eigenvalues for $T$, and $P_1, \ldots, P_r$ the corresponding
spectral projections. On the other hand, if $\Phi(T) = T$, then $T$ belongs to the commutant
$\A^\prime$ by Theorem~\ref{structure}. But again by the Spectral Theorem,
the spectral projections of $T$ commute with
everything that commutes with $T$. Hence $(i) \Rightarrow (ii)$ follows
since $\A^\prime = \fixed $.
\bx

Thus, if we let $\{ P_j \}$ be the (mutually orthogonal) spectral projections of $T$, then
$P_j$ belongs to $\fix(\Phi) = \A^\prime$ for each $j$.
Observe that for all $j$ we may define a map $\Phi_j : \B(P_j \H) \rightarrow \B(P_j \H)$ by
\[
\Phi_j (S) = \sumkn A_{k,j} S A_{k,j}^\dagger,
\]
where
\[
A_{k,j} \equiv A_k P_j = P_j A_k = P_j A_k P_j.
\]
It should be kept in mind that this is a slight abuse of notation; we
should really write $A_{k,j}$
as the restriction $A_{k,j} = A_k|_{P_j\H}$.
These are unital channels since
\[
\Phi_j (\one_{P_j \H}) = \sumkn A_{k,j} A_{k,j}^\dagger = \sumkn P_j A_k A_k^\dagger
P_j
= P_j \Phi(\one_\H) P_j = \one_{P_j \H}.
\]
A similar argument can be used to verify trace preservation for $\Phi_j$; that is, $\sumkn
A_{k,j}^\dagger A_{k,j} = \one_{P_j \H}$.
Proposition~\ref{minimal} below uses an analysis of these maps to give a test for deciding if
$P_j$ is minimal.

\begin{lem}\label{trivialminimal}
Let $\Psi : \bofh \rightarrow \bofh$ be a unital quantum channel, with noise operators $C_1,
\ldots, C_m$, and let $\C = \Alg \{ C_1, \ldots, C_m \}$. The following are equivalent:
\begin{itemize}
\item[$(i)$] $\fix (\Psi) = \C^\prime = \bbC \one_\H$.
\item[$(ii)$] There are no non-trivial reducing subspaces for the algebra $\C$.
\end{itemize}
\end{lem}

\Prf
This immediately follows from Theorem~\ref{structure}, since $\fix (\Psi) = \C^\prime$ and
$\C$ is a $\dagger$-algebra.
\bx

\begin{prop}\label{minimal}
Let $\Psi : \bofh \rightarrow \bofh$ be a unital channel with noise operators $\{C_i\}$, and let
$P\in \bofh$ be a projection which reduces the noise operator algebra $\C$ for $\Psi$. Then the
unital channel  $\Psi_P : \B(P\H) \rightarrow \B(P\H)$ given by
$\Psi_P (S) = \sum_i C_{i,P}\, S\, C_{i,P}^\dagger$, where $C_{i,P} = C_i
P = PC_i$, satisfies
\[
\fix (\Psi_P) = (\C P)^\prime = P \C^\prime P .
\]
Furthermore, $P$ is a minimal projection in $\C^\prime$ if and only if
\[
\fix (\Psi_P) = P \C^\prime P = \bbC P.
\]
\end{prop}

\Prf
Again, the statement $(\C P)^\prime = P \C^\prime P$ is a notational
convenience.
We really mean that the compressed commutant, $ P
\C^\prime|_{P\H}$, is equal to the commutant of the restriction,
$(\C|_{P\H})^\prime = (P \C|_{P\H})^\prime$, inside $\B (P\H)$. To see
this, first let $T$ belong to $\C^\prime$. Then \begin{eqnarray*}
(PTP)(C_iP) &=& P(T C_i) P =P( C_i T)P \\
&=& (PC_iP)(PTP) = (C_i P)(PTP),
\end{eqnarray*}
so that $(\C
P)^\prime$ contains $P\C^\prime P$. Conversely, if $T = PTP$ is in $(\C P)^\prime$, then
\[
TC_i = TPC_i  =T(C_iP)= (C_i P)T = C_i T,
\]
and $T$ belongs to $P\C^\prime P$.
Thus Theorem~\ref{structure} gives us
\[
\fix (\Psi_P) = \big\{ C_{i,P}\big\}^\prime = \big\{ C_i P \big\}^\prime = ( \C  P)^\prime =
P \C^\prime P.
\]
The rest of the result follows from the previous lemma.
\bx

If we apply this result for $P=P_j$, we see that $P_j$ is a
minimal projection inside $\A^\prime$ if and only if $P_j
\A^\prime P_j = \bbC P_j$. Thus, since we are assuming the noise
commutant $\A^\prime$ is known as a linear span, we can simply
compute $P_jBP_j$ for a set of vector space generators $B$ for
$\A^\prime$ and check if each $P_j BP_j$ belongs to $\bbC P_j$. If
this is the case, then $P_j$ is minimal and it belongs to the
maximal family of projections we seek. On the other hand, if there
is a $B_0$ in $\A^\prime$ such that $P_j B_0 P_j$ is not in $\bbC
P_j$, then $P_j$ is not minimal; however, in this case we may
iterate the process outlined above. Indeed, $\Phi_j$ is a unital
channel on $\B(P_j \H)$ with non-scalar fixed points, in fact $P_j
B_0 P_j \in \A^\prime P_j = \fix (\Phi_j)$ is such an operator.
Thus, we may use $P_j B_0 P_j$ to obtain a self-adjoint operator
in $\fix (\Phi_j)$, and continue the process by examining the
spectral projections of this operator, which are, of course, each
a subprojection of $P_j$. As $\H$ is finite dimensional, this
process will eventually terminate. We will be left with the
maximal family (since they sum to the identity operator) of
mutually orthogonal non-zero minimal projections in $\A^\prime$,
and this completes Part I.

\subsection{Part II}\label{sS:partII}

We emphasize the importance of distinguishing the various links between
projections in the maximal family from Part~I by presenting the following simple example.

\begin{eg}
{\rm
The point is that
we must distinguish between bonafide orthogonal direct sums on the one hand and  ampliation
algebras on the other by finding linked projections. Let
\[
\C =
\left\{ \left( \begin{matrix}
a & 0 \\
0 & a
\end{matrix} \right) : a\in\bbC \right\}
\simeq \bbC \one_2 \simeq \bbC \otimes \one_2,
\]
be a scalar ampliation algebra
and let $\D$ be the unlinked orthogonal direct sum
given by
\[
\D =
\left\{ \left( \begin{matrix}
a & 0 \\
0 & b
\end{matrix} \right) : a,b \in\bbC \right\}
\simeq \bbC {\mathbf 1} \oplus \bbC {\mathbf 1},
\]
where both of these $\dagger$-algebras are diagonal with respect to the
orthonormal basis $\{ \ket{0}, \ket{1}\}$
for $\bbC^2$.

The family $\big\{ \ket{0}\bra{0}, \ket{1}\bra{1} \big\}$ of rank
one projections is the maximal family  from Part~I for both
algebras. However, the commutant of the scalar ampliation algebra
$\C = \bbC \one_2$ is unitarily equivalent    to $\C^\prime \simeq
\M_2$, whereas the commutant of the orthogonal direct sum $\D
\simeq \bbC {\mathbf 1}\oplus \bbC {\mathbf 1}$ is also $\D^\prime
\simeq \bbC {\mathbf 1}\oplus \bbC {\mathbf 1}$, the direct sum of
two copies of the one dimensional scalar algebra. Notice that the
projections $\ket{0}\bra{0}$ and $\ket{1}\bra{1}$ are linked in
$\C$, since $\C$ contains $\one_2 = \ket{0}\bra{0} +
\ket{1}\bra{1}$ and neither of these subprojections belongs to
$\C$. On the other hand, they are unlinked in $\D$ because both
projections already belong to $\D$. }
\end{eg}

Before continuing, let us recall a helpful identity from operator
algebras. As a $\dagger$-algebra, $\A$ is equal to its own double
commutant, $\A = (\A^\prime)^\prime= \A^{\prime\prime}$. This is a
special case of von Neumann's double commutant theorem
\cite{Arvinvite,byeg,Tak}. Now, let $\{P_j\}$ be the family of
projections for $\A$ obtained in   Part~I. It follows from the
discussion at the start of this section that links can only occur
between $P_j$ of the same rank. Hence for $k\geq 1$, let
$\{P_{j,k}\}$ be the projections amongst $\{P_j\}$ with $\dim
P_{j,k}\H = k$. Further let $\S_k$ be the index set for this set
of projections.  Suppose $\A^\prime = \spn\{B_1, \ldots, B_r\}$
with $B_i $ in $\bofh$.

Since $\A = (\A^\prime)^\prime = \big( \spn\{B_1, \ldots, B_r\}\big)^\prime$, we may check if
projections of the form $P_\S = \sum_{j\in\S} P_{j,k}$, where $\S\subseteq \S_k$, belong to
$\A$ simply by computing the commutators
\[
[P_\S, B_i ] = P_\S B_i - B_i P_\S \qfor i =1,\ldots, r.
\]
The projections $\{P_{j,k}\}_{j\in\S}$ are linked in $\A$
precisely when $[P_\S,B_i] = 0$ for each $i$, and, there is no
proper subset $\S_1$ of $\S$ such that $[P_{\S_1},B_i] = 0$ for
each $i$.As $\H$ is finite dimensional, there are just finitely
many computations required here. Thus, we will eventually exhaust
all possibilities and discover the links between the projections
$\{P_{j,k}\}$.

Testing for links between the rank one projections in $\{P_j\}$ is easier than the
general case, hence we present it separately. Recall a rank one projection $Q = \ket{\xi}
\bra{\xi}$ belongs to $\A^\prime = \fixed$ if and only if $\xi$ is a joint eigenvector for $A_1,
\ldots, A_n$.

\begin{lem}\label{rankonetest}
Let $\{Q_k \}$ be the rank one projections from the set $\{P_j\}$.
Let $Q_k = \ket{\xi_k}\bra{\xi_k}$ and suppose $\lambda_{ik}$ are scalars with
\[
A_i \xi_k = \lambda_{ik} \xi_k \qfor i=1, \ldots, n.
\]
Define a function $f: \{Q_k\} \longrightarrow \bbC^n$ by
\[
f(Q_k) = (\lambda_{1k}, \ldots, \lambda_{nk}).
\]
If $\lambda$ in $\bbC^n$ is such that $f^{-1}(\{\lambda \})$ is nonempty, then the projections
in  $f^{-1}(\{\lambda \})$ are linked in $\A$.
\end{lem}

\Prf
Let $f^{-1}(\{\lambda \})$ be a nonempty subset of $\{Q_k \}$. Observe that if $\xi_k$ and
$\xi_l$ are range vectors for projections $Q_k$ and $Q_l$ in $f^{-1}(\{\lambda \})$, then they
are eigenvectors for
$\A$ and we have
\begin{eqnarray}\label{rkonelink}
\bra{A\xi_k} \ket{\xi_k} = \bra{A\xi_l} \ket{\xi_l} \qforal A\in\A.
\end{eqnarray}
Indeed, $f(Q_k) = f(Q_l)$ and hence these inner products agree on a set of generators $A_1,
\ldots, A_n$ for $\A$.
Suppose $Q$ is a projection in $\A$ which is the sum of some, but not all, of the projections in
$f^{-1}(\{\lambda \})$. Then there is a pair of vectors $\xi_k$ and $\xi_l$, each a range vector
for a projection in  $f^{-1}(\{\lambda \})$,  such that $Q\xi_k =
\xi_k$ and $Q\xi_l = 0$. In particular, the identity (\ref{rkonelink}) does not hold for $Q$, and
consequently $Q$ does not belong to $\A$.

For a nonempty set $f^{-1}(\{\lambda \})$, let $Q_\lambda$ be the sum of all  projections
inside $f^{-1}(\{\lambda \})$. It remains to show that each $Q_\lambda$ belongs to
$\A$. From the structure theory for  $\dagger$-algebras discussed at the start of this section, we
know that for
each $m$ the projection  $\sum_{j\in \S_m} P_j$ belongs to $\A$, where $\S_m$ is the index
set for all projections in $\{P_j \}$ of rank $m$. Let $P= \sum_{j\in \S_1} P_j =\sum_k
\ket{\xi_k}\bra{\xi_k}$ be the projection
obtained for $m=1$. Then it is easy to see that the compressions $A_i P = P A_i$, for $i
=1,\ldots ,n$, are mutually commuting  operators
inside $\A$ which are each diagonal with respect to the basis $\{ \xi_k\}$ for the range of $P$.
Hence it follows that $\A P$ is a subalgebra of $\A$ which, up to unitary equivalence,
has a block matrix decomposition of the form
\[
\A P \simeq \left( \begin{matrix}
a_{n_1} \one_{n_1} & & 0 \\
  & \ddots &  \\
0 & & a_{n_d} \one_{n_d}
\end{matrix}\right).
\]
Further, under this unitary equivalence, each projection onto a block $\one_{n_j}$ corresponds
to a projection which is the sum of projections from a particular  nonempty $f^{-1}(\{\lambda
\})$. Since complete freedom is allowed for the scalars $a_{n_j}$, we see that every
$Q_\lambda$ belongs to $\A$.
\bx

\begin{rem}
{\rm As a prelude to some of the examples below, we note that the
case of an abelian algebra $\A$ corresponds to the case that the
family of minimal $\A$-reducing projections are all rank one.}
\end{rem}

\subsection{Summary of the Algorithm}\label{sS:summary}

We present a number of applications in the rest of the paper, but let us finish this section with a
brief technical summary of the algorithm. Recall we have assumed that a set of vector space
generators is known for $\fixed = \A^\prime = \spn \{ B_1, \ldots, B_r \} $.

\vspace{0.1in}

{\noindent}{\bf Part I.}

\begin{itemize}
\item[$(i)$] If $\A^\prime \neq \bbC \one$, then choose $T =
T^\dagger$ inside $\A^\prime \setminus \bbC \one$. \item[$(ii)$]
Compute the spectral projections $P_1, \ldots , P_s$ for $T$.
\item[$(iii)$] For each $P = P_j$, compute $PB_1P, \ldots, PB_r
P$. If every $PB_i P$ belongs to $\bbC P$, then $P\A^\prime P =
\bbC P$ and $P$ is a (non-zero) minimal reducing projection for
$\A$, and it belongs to  the maximal family. \item[$(iv)$] If some
$B=B_i$ is such that $PB P$ does not belong to $\bbC P$, then $P$
is not minimal $\A$-reducing and the compression channel $\Phi_P$
has non-scalar fixed points; $PBP$ for instance. Then proceed
through steps $(i)$, $(ii)$ and $(iii)$, perhaps a number of
times, to find minimal $\A$-reducing mutually orthogonal
projections supported on $P$.
\end{itemize}

{\noindent}{\bf Part II.}

\begin{itemize}
\item[$(i)$]
Given the family $\{P_j\}$ from Part I, group together those projections
with the same rank.
\item[$(ii)$]
For the projections $\{ P_{j,k}\}_{j\in\S_k}$ of rank $k\geq 2$, test for links between the
$P_{j,k}$ by evaluating the commutators $[P_\S, B_i]$ where $P_\S = \sum_{j\in\S} P_{j,k}$
with $\S \subseteq \S_k$. Then $\{ P_{j,k} : k \in \S \}$ are linked in $\A$ if and only if
$[P_\S,B_i] = 0$ for
each $i$, and there is no proper subset of $\S$ with this property.
The projections of rank $k=1$ may be tested using Lemma~\ref{rankonetest}. The projections
$P_\S$ defined by linked sets of projections will form the family of minimal central projections
for $\A$.
\item[$(iii)$]
The various links will give the $\dagger$-algebra structure of $\A$ up to unitary equivalence,
and hence of $\A^\prime =
\fixed$, along with the precise locations of matrix blocks inside $\A^\prime$.
\end{itemize}

\section{Phase Damping and Related Examples}\label{S:examples}

In specific cases the computations involved with the method
described above can be less cumbersome than what is suggested. We
begin with a number of simple illustrative applications of the
process. In each of these cases the commutant is already
well-known, or at least easy to find, but we shall demonstrate how
to compute it using the algorithm. See \cite{FVHTC,KBLW} for
related examples.

Recall the Pauli matrices are given by
\[
\one_2 =  \left( \begin{matrix}
1 & 0 \\
0 & 1
\end{matrix}\right), \,\,
X =  \left( \begin{matrix}
0 & 1 \\
1 & 0
\end{matrix}\right), \,\,
Y =  \left( \begin{matrix}
0 & -i \\
i & 0
\end{matrix}\right), \,\,
Z =  \left( \begin{matrix}
1 & 0 \\
0 & -1
\end{matrix}\right).
\]
We  regard these as matrix representations of operators acting on $\bbC^2$ with respect to
an orthonormal  basis $\{ \ket{0}, \ket{1} \}$ corresponding to the base states of a given
two-level system. In the case of higher dimensions, for instance
$\bbC^4$ with  basis $\{ \ket{00}, \ket{01}, \ket{10}, \ket{11}\}$, we use notation
such as $Z_1 $ and $Z_2$ to denote, respectively, the tensor products of $Z \otimes \one_2$
and $ \one_2 \otimes Z$ acting on $\bbC^4$.

\begin{eg}\label{easy1}
{\rm
Let $p$ be a positive real number, $0 < p < 1$. Let $A_1, A_2$ be operators on $\bbC^2$
defined on the standard  basis by:
\[
A_1 = (\sqrt{1-p}) \one_2 = \sqrt{1-p} \left( \begin{matrix}
1 & 0 \\
0 & 1
\end{matrix}\right)
\]
and
\[
A_2 = (\sqrt{p}) Z = \sqrt{p} \left( \begin{matrix}
1 & 0 \\
0 & -1
\end{matrix}\right).
\]
Clearly $\sum_{k=1}^2 A_kA_k^\dagger = \one_2 = \sum_{k=1}^2 A_k^\dagger
A_k$, so
$\{A_1,A_2\}$ are  noise operators for a unital channel $\Phi$
acting on $\B (\bbC^2)$.  The quantum operation corresponding to this  channel is equivalent to
the {\it phase flip} or {\it phase damping} operation on single qubits \cite{NC}. Its effect is to
flip the  phase of $\ket{0}$ relative  to that of  $\ket{1}$, and vice-versa, with probability $p$.

A simple computation shows that
\begin{eqnarray*}
\fixed = \A^\prime &=& \{A_1,A_2\}^\prime \\
&=& \left\{   \left( \begin{matrix}
a & 0 \\
0 & b
\end{matrix}\right) : a,b\in\bbC \right\} \simeq \bbC {\mathbf
1}\oplus \bbC {\mathbf 1} \\
&=& \Alg\{ Z \} = \A.
\end{eqnarray*}
Let us see how this structure arises through the algorithm.
First choose a non-scalar operator in $\A^\prime$; in this case $Z$ will
suffice. The rank one projections $P_0 =
\ket{0} \bra{0}$ and $P_1 = \ket{1} \bra{1}$
are the spectral projections for $Z$, hence they are minimal
inside $\A^\prime$ since the vectors $\ket{0}$ and $\ket{1}$
are joint eigenvectors for $A_1,A_2$.
Thus the family $\{ P_0 , P_1\}$ forms
the maximal family ($\one_2 = P_0 + P_1$)
of non-zero mutually orthogonal minimal projections inside the commutant $\A^\prime$.

Further observe that
\[
\left\{ \begin{array}{lcl}
A_1 \ket{0} = \sqrt{1-p}\,\, \ket{0} \\
A_2 \ket{0} = \sqrt{p}\,\,\ket{0}
\end{array}\right.
\qand
\left\{ \begin{array}{lcl}
A_1 \ket{1} = \sqrt{1-p}\,\, \ket{1} \\
A_2 \ket{1} = - \sqrt{p} \,\, \ket{1}
\end{array}\right.
\]
The 2-tuples $(\sqrt{1-p},\sqrt{p})$ and $(\sqrt{1-p},- \sqrt{p})$
are distinct, so the projections $P_0$ and $P_1$ are {\it not}
linked by Lemma~\ref{rankonetest}. Hence we have $\A = \A P_0
\oplus \A P_1$, with $\A P_0$ and $ \A P_1$ both subalgebras of
$\A$ each unitarily equivalent to the one dimensional scalar
algebra $\bbC {\mathbf 1}$. Thus, the commutant $\A^\prime = (\A
P_0)^\prime + (\A P_1)^\prime  = P_0 \A^\prime P_0 + P_1 \A^\prime
P_1 $ is also unitarily equivalent to $\bbC {\mathbf 1}\oplus \bbC
{\mathbf 1}$, and the form of $\fixed = \A^\prime$ given above is
evident. }
\end{eg}

\begin{eg}\label{easy2}
{\rm
Let $p$ be a positive real number, $0 < p < 1$. Let $A_1, A_2$ be operators on $\bbC^4$
defined on the standard orthonormal basis by:
\[
A_1 = \sqrt{1-p} \,\, (\one_2 \otimes \one_2) =  \sqrt{1-p} \left( \begin{matrix}
1 & 0 & 0 & 0 \\
0 & 1 & 0 & 0 \\
0 & 0 & 1 & 0 \\
0 & 0 & 0 & 1
\end{matrix}\right),
\]
and
\[
A_2 = \sqrt{p}\,\, (Z_1  Z_2) =\sqrt{p}\,\, (Z \otimes  Z)  =  \sqrt{p} \left( \begin{matrix}
1 & 0 & 0 & 0 \\
0 & -1 & 0 & 0 \\
0 & 0 & -1 & 0 \\
0 & 0 & 0 & 1
\end{matrix}\right),
\]
Once again,  it is clear that these noise operators determine a unital
channel $\Phi$ on $\B (\bbC^4)$.
We may compute
\begin{eqnarray*}
\fixed = \A^\prime &=& \{A_1,A_2\}^\prime  \\
&=&  \Alg\{ Z_1, Z_2, X_1 X_2 + Y_1 Y_2 \} \\
&=& \left\{   \left( \begin{matrix}
a_{11} & 0 & 0 & a_{14} \\
0 & a_{22} & a_{23} & 0  \\
0 & a_{32} & a_{33} & 0  \\
a_{41} & 0 & 0 & a_{44}
\end{matrix}\right) : a_{ij} \in\bbC \right\}.
\end{eqnarray*}

By observation, we can see that $\A^\prime$ is unitarily
equivalent to the orthogonal direct sum $\A^\prime \simeq \M_2
\oplus \M_2$. To see this through the algorithm,  first choose a
non-scalar self-adjoint operator in $\fixed = \A^\prime$;
\[
Z_1 = Z\otimes \one_2 = \left( \begin{matrix}
1 & 0 & 0 & 0 \\
0 & 1 & 0 & 0  \\
0 & 0 & -1 & 0  \\
0 & 0 & 0 & -1
\end{matrix}\right)
= Z_1^\dagger
\]
is such an operator. The spectral projections for $Z_1$, corresponding to eigenvalues $\lambda
= 1$ and $\lambda = -1$, are given by
\[
P_1 = \ket{00}\bra{00} + \ket{01}\bra{01} \qand
P_{-1} = \ket{10}\bra{10} + \ket{11}\bra{11}.
\]
Let $P_{kj} = \ket{kj}\bra{kj} $ for $k,j \in \{0,1\}$. The projections  $P_1$ and $P_{-1}$
belong to $\A^\prime$, but in this case they are not minimal inside $\A^\prime$. Indeed, observe
that
\begin{eqnarray*}
P_1 \A^\prime P_1 &=& (\A P_1 )^\prime  =  \{A_1 P_1, A_2 P_1 \}^\prime \\
& = & \left\{   \left( \begin{matrix}
a_{11} & 0 & 0 & 0 \\
0 & a_{22} & 0 & 0  \\
0 & 0 & 0 & 0  \\
0 & 0 & 0 & 0
\end{matrix}\right) : a_{11},a_{22} \in\bbC \right\}
\neq \bbC P_1,
\end{eqnarray*}
and similarly for $P_{-1}$. Following the algorithm, we let
\[
Z_1 Z_2 P_1 = P_1 Z_1 Z_2 =
\left( \begin{matrix}
1 & 0 & 0 & 0 \\
0 & -1 & 0 & 0  \\
0 & 0 & 0 & 0  \\
0 & 0 & 0 & 0
\end{matrix}\right)
\]
be a
non-scalar operator in $P_1 \A^\prime P_1 $. The spectral projections for this operator are
\[
P_{00} = \ket{00}\bra{00}
\qand
P_{01} = \ket{01}\bra{01}.
\]
These rank one projections are minimal inside $\A^\prime$. Similarly, we find the
subprojections $P_{10} = \ket{10}\bra{10}$ and $P_{11} = \ket{11}\bra{11}$ of $P_{-1}$
are minimal inside $\A^\prime$, and hence the maximal family of non-zero mutually orthogonal
minimal projections in $\A^\prime$ is given by
\[
\big\{ P_{00},  P_{01}, P_{10}, P_{11} \big\} .
\]

It remains to check for links between these minimal projections. Observe that
\[
\left\{ \begin{array}{lcl}
A_1 \ket{00} = \sqrt{1-p}\,\, \ket{00} \\
A_2 \ket{00} = \sqrt{p}\,\,\ket{00}
\end{array}\right.
\qquad
\left\{ \begin{array}{lcl}
A_1 \ket{01} = \sqrt{1-p}\,\, \ket{01} \\
A_2 \ket{01} = - \sqrt{p}\,\,\ket{01}
\end{array}\right.
\]
\[
\left\{ \begin{array}{lcl}
A_1 \ket{10} = \sqrt{1-p}\,\, \ket{10} \\
A_2 \ket{10} = - \sqrt{p}\,\,\ket{10}
\end{array}\right.
\qquad
\left\{ \begin{array}{lcl}
A_1 \ket{11} = \sqrt{1-p}\,\, \ket{11} \\
A_2 \ket{11} =  \sqrt{p}\,\,\ket{11}
\end{array}\right.
\]
Thus, upon comparing the corresponding 2-tuples $(\sqrt{1-p},
\sqrt{p})$ and $(\sqrt{1-p}, - \sqrt{p})$, we see that the pairs
$\{ P_{00}, P_{11}\}$ and $\{ P_{01}, P_{10}\}$ are linked in $\A$
by Lemma~\ref{rankonetest}. In particular, $P_{00}+ P_{11}$ and
$P_{01}+ P_{10}$ belong to $\A = \A ( P_{00}+ P_{11}) \oplus \A (
P_{01}+ P_{10})$ and $\A$ is unitarily equivalent to $\A \simeq
\bbC \one_2 \oplus \bbC \one_2$.

Hence, $\fixed = \A^\prime \simeq \M_2 \oplus \M_2$, and the
locations of the $\M_2$ blocks inside $\fixed = \A^\prime$ are
explicitly given by the algorithm. Namely, one block of $2\times
2$ matrices inside $\A^\prime = \fixed$ is given by all $T$ in
$\B(\bbC^4)$ such that $T = (P_{00}+P_{11} )T (P_{00}+P_{11} ),$
and similarly the other $\M_2$ block is supported on $P_{01} +
P_{10}$. The blocks are disjoint in $\A^\prime$ in the sense that
if $T$ belongs to $\M_2$, then an operator $\td{T} =
(P_{00}+P_{11} )T (P_{00}+P_{11} )$ can be defined in $\A^\prime$,
and similarly for the $P_{01} + P_{10}$ block. }
\end{eg}

\begin{eg}\label{easy3}
{\rm
Let $p$ be a positive real number, $0 < p < 1$.  Let $ A_1, A_2, A_3,A_4$ be operators on
$\bbC^4$ defined on the standard  basis by:
\[
A_1 = (1-p) \,\, (\one_2 \otimes \one_2) = (1-p) \left( \begin{matrix}
1 & 0 & 0 & 0 \\
0 & 1 & 0 & 0 \\
0 & 0 & 1 & 0 \\
0 & 0 & 0 & 1
\end{matrix}\right)
\]
\[
A_2 = (\sqrt{p(1-p)})\,\, Z_1  =  \sqrt{p(1-p)} \left( \begin{matrix}
1 & 0 & 0 & 0 \\
0 & 1 & 0 & 0 \\
0 & 0 & -1 & 0 \\
0 & 0 & 0 & -1
\end{matrix}\right)
\]
\[
A_3 = (\sqrt{p(1-p)})\,\, Z_2 =  \sqrt{p(1-p)} \left( \begin{matrix}
1 & 0 & 0 & 0 \\
0 & -1 & 0 & 0 \\
0 & 0 & 1 & 0 \\
0 & 0 & 0 & -1
\end{matrix}\right)
\]
\[
A_4 = (p) \,\, Z_1  Z_2  = (p) Z\otimes Z =  p \left( \begin{matrix}
1 & 0 & 0 & 0 \\
0 & -1 & 0 & 0 \\
0 & 0 & -1 & 0 \\
0 & 0 & 0 & 1
\end{matrix}\right)
\]
These noise operators determine a unital channel $\Phi$ on $\B(\bbC^4)$.
The fixed point algebra may be computed as
\begin{eqnarray*}
\fixed = \A^\prime &=& \{A_1,A_2, A_3, A_4 \}^\prime =   \\
&=& \left\{   \left( \begin{matrix}
a & 0 & 0 & 0 \\
0 & b & 0 & 0  \\
0 & 0 & c & 0  \\
0 & 0 & 0 & d
\end{matrix}\right) : a,b,c,d \in\bbC \right\} \\
&\simeq& \bbC {\mathbf 1}\oplus \bbC {\mathbf 1} \oplus \bbC
{\mathbf 1}\oplus \bbC {\mathbf 1}
\end{eqnarray*}

Running through the algorithm, we first choose $Z_1 = Z\otimes \one_2 = Z_1^\dagger$ as a
non-scalar operator in $\A^\prime$. As in the previous example, the spectral projections for
$Z_1$ are given by $P_1$ and $P_{-1}$. Once again, neither of these projections is minimal
inside $\A^\prime$. For instance, for $ P_1 = \ket{00}\bra{00} + \ket{01}\bra{01}$ the fixed
point set of the restricted channel $\Phi_{P_1} (S) = \sum_{i=1}^4 A_{k,1} S
A_{k,1}^\dagger$,
where $A_{k,1} = A_k P_1$, satisfies
\begin{eqnarray*}
\fix(\Phi_{P_1}) &=& ( \A P_1)^\prime = P_1 \A^\prime P_1 \\
&=&
\left\{   \left( \begin{matrix}
a & 0 & 0 & 0 \\
0 & b & 0 & 0  \\
0 & 0 & 0 & 0  \\
0 & 0 & 0 & 0
\end{matrix}\right) : a,b \in\bbC \right\} \neq \bbC P_1,
\end{eqnarray*}
and hence contains non-scalar operators.

It follows that the projections
$
\big\{ P_{00},  P_{01}, P_{10}, P_{11} \big\}
$
again form the maximal family of non-zero mutually orthogonal minimal projections in $\fixed
= \A^\prime$.
But unlike the previous example, there are no links between these rank one projections. Indeed,
observe that the numeric 4-tuples given by the equations
\[
\left\{ \begin{array}{lcl}
A_1 \ket{00} = (1-p)\,\, \ket{00} \\
A_2 \ket{00} = \sqrt{p(1-p)}\,\,\ket{00} \\
A_3 \ket{00} = \sqrt{p(1-p)}\,\,\ket{00} \\
A_4 \ket{00} = p \,\,\ket{00}
\end{array}\right.
\qquad
\left\{ \begin{array}{lcl}
A_1 \ket{01} = (1-p)\,\, \ket{01} \\
A_2 \ket{01} = \sqrt{p(1-p)}\,\,\ket{01} \\
A_3 \ket{01} = - \sqrt{p(1-p)}\,\,\ket{01} \\
A_4 \ket{01} = -p \,\,\ket{01}
\end{array}\right.
\]
\[
\left\{ \begin{array}{lcl}
A_1 \ket{10} = (1-p)\,\, \ket{10} \\
A_2 \ket{10} = - \sqrt{p(1-p)}\,\,\ket{10} \\
A_3 \ket{10} = \sqrt{p(1-p)}\,\,\ket{10} \\
A_4 \ket{10} = - p \,\,\ket{10}
\end{array}\right.
\qquad
\left\{ \begin{array}{lcl}
A_1 \ket{11} = (1-p)\,\, \ket{11} \\
A_2 \ket{11} = - \sqrt{p(1-p)}\,\,\ket{11} \\
A_3 \ket{11} = - \sqrt{p(1-p)}\,\,\ket{11} \\
A_4 \ket{11} = p \,\,\ket{11}
\end{array}\right.
\]
are pairwise distinct. Thus each projection $P_{ij}$ belongs to
$\A = \A P_{00} + \A P_{01} + \A P_{10} + \A P_{11}$, and $\A$ is
unitarily equivalent to the unlinked orthogonal direct sum $\A
\simeq \bbC {\mathbf 1} \oplus   \bbC {\mathbf 1} \oplus \bbC
{\mathbf 1} \oplus \bbC {\mathbf 1} $. Hence, $\fixed = \A^\prime
\simeq \bbC {\mathbf 1} \oplus   \bbC {\mathbf 1} \oplus \bbC
{\mathbf 1} \oplus \bbC {\mathbf 1} $ as well. }
\end{eg}

\section{Unitization of Channels}\label{S:unitization}

Before proceeding  we present a discussion which will be
useful in the sequel.
Recall from  the Spectral Theorem that every normal operator $T$ in $\B(\H)$ decomposes
as a sum $T = \lambda_1 P_1 + \ldots + \lambda_r P_r$ where $\lambda_i$ are the distinct
eigenvalues for $T$ and $P_i$ the corresponding spectral projections. The {\it functional
calculus} for $T$ is determined by all
complex-valued
functions $f$ which are defined on the eigenvalues of $T$. For each such function, there is an
operator $f(T)$ in the functional calculus for $T$ defined by $f(T) = f(\lambda_1) P_1 +
\ldots + f(\lambda_r) P_r$.

\begin{lem}\label{unitizationlemma}
Let $T$ be a normal operator, and suppose $f$ is a function such that $f(T)$ belongs to the
functional calculus for $T$. If $f$ is
injective on the set of eigenvalues for $T$, then the commutant  $\{T \}^\prime$
coincides with the commutant  $\{f(T)\}^\prime$.
\end{lem}

\Prf
Suppose  $T = \lambda_1 P_1 + \ldots + \lambda_r P_r$
is the spectral decomposition of $T$. Then, by the
Spectral Theorem, the commutant
of $T$ may also be realized as the commutant of its spectral projections;
$\{T \}^\prime = \{ P_1, \ldots, P_r\}^\prime$. By hypothesis, $f(\lambda_i
) = f(\lambda_j)$ implies $i=j$. Thus, it follows  that the spectral
projections for $f(T) = f(\lambda_1) P_1 + \ldots + f(\lambda_r)P_r$ are
precisely $\{P_1, \ldots, P_r\}$. Hence, $\{T\}^\prime = \{
f(T)\}^\prime$, as claimed.
\bx

As  an immediate consequence of Theorem~\ref{structure}, we obtain the following result for
general quantum channels.

\begin{cor}\label{unitization}
If   $\{A_1, \ldots, A_n\}$ are normal operators, and if, for each $k$, $f(A_k)$ is in the
functional calculus of $A_k$ and is injective on the set of eigenvalues for  $A_k$, then
\[
\{A_1, \ldots, A_n\}^\prime = \{f(A_1), \ldots, f(A_n)\}^\prime.
\]
In  particular, if
\[
\Phi (X) = \sumkn f(A_k) X f(A_k)^\dagger
\]
defines a unital quantum channel, then
\[
\fix (\Phi) = \{ A_1, \ldots, A_n\}^\prime.
\]
\end{cor}

Hence this gives a way of `unitizing' a quantum channel, when the
noise operators satisfy the hypotheses of
Corollary~\ref{unitization}, in such a way that the noise
commutant of the new unital channel is completely  determined by
the commutant of the original noise operators. We shall make use
of this lemma in the following section with the function $f$ given
by a multiple of the natural exponential function.

\section{Collective Noise Channels}\label{S:collective}

We next consider a class of examples which are  more involved
technically. The 3-qubit example from this class   has been
analyzed previously in the literature
\cite{Lnoiseless,KBLW,KLV,VKL} and also experimentally realized
via {\it nuclear magnetic resonance} \cite{VFPKLC}. In particular,
the algorithm allows us to explicitly identify the structure of
the noise commutant for this channel. We will also conduct this
analysis on the 4-qubit case, but first we outline the general
$n$-qubit case.

Let $n \geq 3$ be a fixed positive integer.   Let $\mathbf{X}$ be defined on
$\bbC^{2^n}$ by
\[
\mathbf{X} = X_1 + X_2 + \ldots + X_n,
\]
where
\[
X_1 = X \otimes \big( \one_2^{\otimes(n-1)}\big), \,\,\, X_2 = \one_2 \otimes X \otimes \big(
\one_2^{\otimes(n-2)}\big), \ldots
\]
Similarly define $\mathbf{Y} = Y_1+\ldots +Y_n$ and $\mathbf{Z} = Z_1 + \ldots +Z_n$.
Observe that all of these
operators are self-adjoint. Let $\exp (\cdot)$ be the complex-valued
natural exponential function. This function belongs to the functional calculus of each of
$\mathbf{X},\mathbf{Y},\mathbf{Z}$, and hence we may define operators $E_x , E_y, E_z$ on
$\bbC^{2^n}$ by
\[
E_x = \frac{1}{\sqrt{3}} \exp (i \mathbf{X}) , \,\,\,\,\, E_y = \frac{1}{\sqrt{3}} \exp (i
\mathbf{Y})
, \,\,\,\,\,
E_z = \frac{1}{\sqrt{3}} \exp (i \mathbf{Z}) .
\]
A unital channel $\Phi$, called the {\it $n$-qubit collective
noise (rotation) channel}, is then defined on $\B (\bbC^{2^n})$ by
\begin{eqnarray}\label{nqubitdefn}
\Phi (T) = E_x T E_x^\dagger + E_y T E_y^\dagger + E_z T E_z^\dagger.
\end{eqnarray}

As before, let $\A = \Alg \{E_x, E_y, E_z\}$ be the algebra generated by the noise operators.
The following result provides a useful computational device for this class
of channels.

\begin{prop}\label{nqubitcomm}
If $\Phi$ is given by $(\ref{nqubitdefn})$, then
\begin{eqnarray*}
\fixed = \A^\prime &=& \{E_x, E_y, E_x\}^\prime
= \{\mathbf{X},\mathbf{Y},\mathbf{Z}\}^\prime \\
&=& \{\mathbf{X},\mathbf{Y}\}^\prime = \{\mathbf{X},\mathbf{Z}\}^\prime
= \{\mathbf{Y},\mathbf{Z}\}^\prime.
\end{eqnarray*}
\end{prop}

\Prf
The identities in the first line of this equation are a consequence of Theorem~\ref{structure} and
Corollary~\ref{unitization}; for the equality $\{E_x, E_y, E_x\}^\prime
= \{\mathbf{X},\mathbf{Y},\mathbf{Z}\}^\prime$ we use the fact that $\exp (i \cdot)$ is
injective on the set of eigenvalues for each of $\mathbf{X},\mathbf{Y},\mathbf{Z}$. This can
be seen by noting that the spectrum of $\mathbf{Z}$ consists of integers and that
$\mathbf{X},\mathbf{Y},\mathbf{Z}$ are unitarily equivalent.

The equation $\{\mathbf{X},\mathbf{Y},\mathbf{Z}\}^\prime =
\{\mathbf{X},\mathbf{Y}\}^\prime$ follows from the  anti-commutation relations for the Pauli
matrices since
\[
\mathbf{X} \mathbf{Y} - \mathbf{Y} \mathbf{X} = \left( i \mathbf{Z} + \sum_{k\neq j} X_k
Y_j \right) -  \left( -i \mathbf{Z} + \sum_{k\neq j}  Y_j X_k \right) = 2i  \mathbf{Z} .
\]
The other equalities  are similar.
\bx

\begin{rem}
{\rm Notice that we could also define a channel $\Phi$ for $n=1$
and $n=2$. But simple computations in both cases show that $\fixed
= \{ \mathbf{X},\mathbf{Y},\mathbf{Z}\}^\prime$ contains no
bonafide matrix blocks; $\fixed = \bbC \one_2$ for $n=1$, and
$\fixed \simeq \bbC \one_3 \oplus \bbC {\mathbf 1}$ for $n=2$.
Hence these two channels are not of interest from a quantum
computing perspective. On the other hand, there is rich structure
in the commutant  for $n\geq 3$. }
\end{rem}

\section{The 3-Qubit Case}\label{S:3qubitcase}

In this section we consider the $n=3$ case for the collective noise channels.
The matrix representations we use in this
section are with respect to the standard ordered orthonormal basis for $\bbC^8$ associated with
the usual
tensor product notation $A\otimes B = (a_{ij}B)_{ij}$ for matrices. This is the basis
\[
\big\{ \ket{000}, \ket{001}, \ket{010}, \ket{011}, \ket{100}, \ket{101}, \ket{110}, \ket{111}
\big\}.
\]

We show explicitly that the algebra $\A$ is unitarily equivalent
to the unlinked orthogonal direct sum of the full matrix algebra
$\M_4$ together with the ampliation algebra $\M_2 \otimes \one_2$;
\[
\A \simeq \big( \M_2 \otimes \one_2 \big) \oplus \M_4 .
\]
Thus the noise commutant contains an ampliated copy of $\M_2$;
\begin{eqnarray*}
\fixed = \A^\prime &\simeq& \big( \M_2 \otimes \one_2
\big)^\prime \oplus \big( \M_4 \big)^\prime \\
&\simeq& \big( \one_2 \otimes \M_2 \big) \oplus \bbC \one_4 \\
&\simeq&
\left\{   \left( \begin{matrix}
A &  &  &   &  &  \\
 & A &  &  & 0 &  \\
 &  & b &  &  &  \\
  &  &  & b &  &  \\
 & 0 &  &  & b &  \\
 &  &  &  &  & b
\end{matrix}\right) : A\in \M_2, b \in\bbC \right\}
\end{eqnarray*}
We will verify this by using the algorithm to produce the maximal
family of non-zero minimal $\A$-reducing projections $\{ P,Q,R\}$,
and show that $P,Q$ are linked inside $\A$ with $\rank P = 2 =
\rank Q$ and $\rank R = 4$. In particular, this will imply that
$\A = \A (P+Q) \oplus \A R$ with
\[
\M_2 \otimes \one_2 \simeq \A(P+Q) \subseteq \A \qand
\M_4 \simeq \A R \subseteq \A,
\]
and thus $\fixed = \A^\prime = \big( \A (P+Q) \big)^\prime \oplus
\big( \A R \big)^\prime$ with
\[
\M_2 \otimes \one_2 \simeq \big( \A(P+Q)\big)^\prime \subseteq \A^\prime \qand
\bbC \one_4 \simeq \big( \A R\big)^\prime \subseteq \A^\prime.
\]
We will also explicitly identify a copy of the Pauli matrices sitting inside the ampliation $\M_2
\otimes \one_2$ contained in $\A^\prime = \fixed$.

We begin by computing $\A^\prime$ as a linear span. By Proposition~\ref{nqubitcomm} we
have
$
\A^\prime =
\{\mathbf{X},\mathbf{Z}\}^\prime.
$
If we let $B = (b_{ij})_{1\leq i,j \leq 8}$ and consider the commutators $[B, \mathbf{Z}]$ and
$[B, \mathbf{X}]$, we find a system of linear equations determined by the $b_{ij}$ which is
satisfied precisely when $B$ belongs to $\A^\prime$. Specifically, the system includes 30
equations with a total of  20 unknowns. Thus we may apply basic linear algebra (we use Matlab
to put the corresponding matrix in its reduced row echelon form) to obtain a set of vector space
generators for $\A^\prime$. As it turns out, there are five parameters which yield matrices $B_r,
B_s, B_t, B_u, B_v$. That is,
\[
\A^\prime = \spn \{ B_r, B_s, B_t, B_u, B_v\}.
\]
But $B_v = B_u^\dagger$. Hence it is clear from Parts I and II that, for the purposes of the
algorithm, we may restrict our attention to the set $\{ B_r, B_s, B_t, B_u\}$. The matrix
representations of these  operators with respect to the standard basis for $\bbC^8$  are given
below.
\[
B_r = B_r^\dagger =
 \left( \begin{matrix}
1 & 0 & 0 & 0 & 0& 0&0 &0 \\
0 & 0 & 0 & 0 & 1& 0&0 &0 \\
0 & 0 & 0 & 0 & 1& 0&0 &0 \\
0 & 0 & 0 & -1 & 0& 1&1 &0 \\
0 & 1 & 1 & 0 & -1& 0&0 &0 \\
0 & 0 & 0 & 1 & 0 & 0&0 &0 \\
 0 & 0 & 0 & 1 & 0 & 0&0 &0 \\
0 & 0 & 0 & 0 & 0 & 0&0 &1
\end{matrix}\right)
\]
\[
B_s =
  \left( \begin{matrix}
0 & 0 & 0 & 0 & 0& 0&0 &0 \\
0 & 0 & 0 & 0 & 0& 0&0 &0 \\
0 & 0 & 1 & 0 & -1& 0&0 &0 \\
0 & 0 & 0 & 1 & 0& -1&0 &0 \\
0 & 0 & -1 & 0 & 1& 0&0 &0 \\
0 & 0 & 0 & -1 & 0 & 1&0 &0 \\
 0 & 0 & 0 & 0 & 0 & 0&0 &0 \\
0 & 0 & 0 & 0 & 0 & 0&0 &0
\end{matrix}\right)
\]
\[
B_t = B_t^\dagger =
 \left( \begin{matrix}
0 & 0 & 0 & 0 & 0& 0&0 &0 \\
0 & 1 & 0 & 0 & -1& 0&0 &0 \\
0 & 0 & 0 & 0 & 0& 0&0 &0 \\
0 & 0 & 0 & 1 & 0& 0&-1 &0 \\
0 & -1 & 0 & 0 & 1& 0&0 &0 \\
0 & 0 & 0 & 0 & 0 & 0&0 &0 \\
 0 & 0 & 0 & -1 & 0 & 0&1 &0 \\
0 & 0 & 0 & 0 & 0 & 0&0 &0
\end{matrix}\right)
\]
\[
B_u =
 \left( \begin{matrix}
0 & 0 & 0 & 0 & 0& 0&0 &0 \\
0 & 0 & 1 & 0 & -1& 0&0 &0 \\
0 & 0 & 0 & 0 & 0& 0&0 &0 \\
0 & 0 & 0 & 1 & 0& -1&0 &0 \\
0 & 0 & -1 & 0 & 1& 0&0 &0 \\
0 & 0 & 0 & 0 & 0 & 0&0 &0 \\
 0 & 0 & 0 & -1 & 0 & 1&0 &0 \\
0 & 0 & 0 & 0 & 0 & 0&0 &0
\end{matrix}\right)
\]

For Part I of the algorithm we shall conduct the spectral analysis on $B_r = B_r^\dagger$. The
characteristic polynomial of $B_r$ may be computed as $\det (\lambda \one_8 - B_r) =
\lambda^2 (\lambda - 1)^4 (\lambda + 2)^2$. Let $P_0, P_1, P_{-2}$ be the projections onto the
eigenspaces for the eigenvalues $\lambda = 0$, $\lambda = 1$, and $\lambda = -2$. These
projections belong to $\A^\prime$ with $\rank P_0 = 2 = \rank P_{-2}$ and $\rank P_1 = 4$.
Define vectors $\{ \xi_0, \eta_0, \xi_{-2}, \eta_{-2} \}$ in $\bbC^8$ as follows:
\[
\xi_0 = \frac{1}{\sqrt{2}}
 \left( \begin{matrix}
0 \\
1 \\
-1 \\
0 \\
0 \\
0 \\
0 \\
0
\end{matrix}\right)
\eta_0 = \frac{1}{\sqrt{2}}
 \left( \begin{matrix}
0 \\
0 \\
0 \\
0 \\
0 \\
1 \\
-1 \\
0
\end{matrix}\right)
\xi_{-2} = \frac{1}{\sqrt{6}}
 \left( \begin{matrix}
0 \\
1 \\
1 \\
0 \\
-2 \\
0 \\
0 \\
0
\end{matrix}\right)
\eta_{-2} = \frac{1}{\sqrt{6}}
 \left( \begin{matrix}
0 \\
0 \\
0 \\
2 \\
0 \\
-1 \\
-1 \\
0
\end{matrix}\right).
\]
We note that these vectors are essentially the `singlet-triplet'
basis for the noiseless representation as given in \cite{VKL} and
also used in \cite{DBKBW,Lnoiseless,KBLW}. Then $P_0$ and $P_{-2}$
are given by
\begin{eqnarray*}
P_0 &=& \ket{\xi_0}\bra{\xi_0} + \ket{\eta_0}\bra{\eta_0} \\
&=&
  \left( \begin{matrix}
0 & 0 & 0 & 0 & 0& 0&0 &0 \\
0 & 1 / 2 & -1 / 2 & 0 & 0& 0&0 &0 \\
0 & -1 / 2 & 1 / 2 & 0 & 0& 0&0 &0 \\
0 & 0 & 0 & 0 & 0& 0 &0 &0 \\
0 & 0 & 0 & 0 & 0& 0&0 &0 \\
0 & 0 & 0 & 0 & 0 & 1 / 2& -1 / 2 &0 \\
 0 & 0 & 0 & 0 & 0 & -1 / 2& 1 / 2 &0 \\
0 & 0 & 0 & 0 & 0 & 0&0 &0
\end{matrix}\right) ,
\end{eqnarray*}
and
\begin{eqnarray*}
P_{-2} &=& \ket{\xi_{-2}}\bra{\xi_{-2}} + \ket{\eta_{-2}}\bra{\eta_{-2}} \\
&=&
  \left( \begin{matrix}
0 & 0 & 0 & 0 & 0& 0&0 &0 \\
0 & 1 / 6 & 1 / 6 & 0 & -1 / 3& 0&0 &0 \\
0 & 1 / 6 & 1 / 6 & 0 & -1 / 3& 0&0 &0 \\
0 & 0 & 0 & 2 / 3 & 0& -1 / 3 &-1 / 3 &0 \\
0 & -1 / 3 & -1 / 3 & 0 & 2 / 3& 0&0 &0 \\
0 & 0 & 0 & -1 / 3 & 0 & 1 / 6& 1 / 6 &0 \\
 0 & 0 & 0 & -1 / 3 & 0 & 1 / 6& 1 / 6 &0 \\
0 & 0 & 0 & 0 & 0 & 0&0 &0
\end{matrix}\right).
\end{eqnarray*}
Furthermore, $P_1$ is computed as
 \begin{eqnarray*}
P_{1} &=& \one_8 - P_0 - P_{-2} \\
&=&
  \left( \begin{matrix}
1 & 0 & 0 & 0 & 0& 0&0 &0 \\
0 & 1 / 3 & 1 / 3 & 0 & 1 / 3& 0&0 &0 \\
0 & 1 / 3 & 1 / 3 & 0 & 1 / 3& 0&0 &0 \\
0 & 0 & 0 & 1 / 3 & 0& 1 / 3 &1 / 3 &0 \\
0 & 1 / 3 & 1 / 3 & 0 & 1 / 3& 0&0 &0 \\
0 & 0 & 0 & 1 / 3 & 0 & 1 / 3& 1 / 3 &0 \\
 0 & 0 & 0 & 1 / 3 & 0 & 1 / 3& 1 / 3 &0 \\
0 & 0 & 0 & 0 & 0 & 0&0 &1
\end{matrix}\right).
\end{eqnarray*}

Next we must examine the compressions of the commutant by these projections. First observe
that $P_\lambda B_r = B_r P_\lambda = \lambda P_\lambda$ for $\lambda = -2, 0 , 1$ as the
$P_\lambda$ are spectral projections for $B_r$. We further compute
\[
P_{-2}B_s P_{-2} = P_{-2}B_t P_{-2} = P_{-2}B_u P_{-2} = \left( \frac{3}{2}\right) P_{-2},
\]
\[
P_{1}B_s P_{1} = P_{1}B_t P_{1} = P_{1}B_u P_{1} = 0,
\]
\[
P_{0}B_s P_{0} = P_{0}B_t P_{0} = \left( \frac{1}{2}\right) P_{0} \qand  P_{0}B_u P_{0} =
     \left(  - \frac{1}{2} \right) P_{0}.
\]
Hence it follows that $P_\lambda \A^\prime P_\lambda = \bbC P_\lambda$ for $\lambda = -2, 0
, 1$. Therefore, the minimal $\A$-reducing projections $\{ P_{-2}, P_0, P_1\}$ form the desired
set of projections.

It remains to check for links. Since linked projections have the same rank, the only possible link
between these projections is $P_{-2} + P_0$. But notice that $P_0 B_u \neq B_u P_0$, and
hence
\[
P_0 \notin \big( \A^\prime \big)^\prime = \A^{\prime\prime} = \A.
\]
Thus, it follows that $P_0$ and $P_{-2}$ must be linked inside $\A$. (One
may check that $P_0
+ P_{-2}\in\A = \A^{\prime\prime}$.) This shows that $\A$, and hence $\A^\prime$, has the
form claimed above.

Finally, we shall exhibit a copy of the Pauli matrices inside $\M_2 \otimes \one_2 \simeq
(P_0 + P_{-2}) \A^\prime (P_0 + P_{-2}) \subseteq \A^\prime = \fixed$. We may accomplish
this by first identifying a set of $2\times 2$ matrix units $\{ E_{11}, E_{12}, E_{21},
E_{22}\}$ inside $(P_0 + P_{-2}) \A^\prime
(P_0 + P_{-2})$. For this
the algebra structure of $\A$ and $\A^\prime$ may be used together
with a spectral analysis of the operators $\{\mathbf{X},\mathbf{Y},\mathbf{Z}\}$.
Specifically,
with the orthonormal basis $\{ \xi_0, \eta_0, \xi_{-2}, \eta_{-2} \}$ for $P_0 \bbC^8 \oplus
P_{-2} \bbC^8$
we may define
\begin{align*}
E_{11} &= \ket{\xi_0} \bra{\xi_0} + \ket{\eta_0} \bra{\eta_0} = P_0 \\
 E_{12} &= \ket{\xi_0} \bra{\xi_{-2}} + \ket{\eta_0} \bra{\eta_{-2}} \\
E_{21} &= \ket{\xi_{-2}} \bra{\xi_0} + \ket{\eta_{-2}} \bra{\eta_0} \\
 E_{22} &= \ket{\xi_{-2}} \bra{\xi_{-2}} + \ket{\eta_{-2}} \bra{\eta_{-2}} = P_{-2}
\end{align*}
Then the matrix representations of operators inside $(P_0 + P_{-2}) \A^\prime (P_0 + P_{-2})$
with respect to the basis $\{ \xi_0,  \xi_{-2}, \eta_0, \eta_{-2} \}$
(a `canonical shuffle'
\cite{Paulsentext,Paulsentext2} of the above basis)
have the form
\[
(P_0 + P_{-2}) \A^\prime (P_0 + P_{-2}) \simeq
\left\{ \left( \begin{matrix}
A & 0 \\
0 & A
\end{matrix} \right) : A\in\M_2 \right\},
\]
and the $E_{ij}$ are the natural matrix units in this decomposition. Thus, we may obtain
versions of the Pauli matrices inside $\A^\prime = \fixed$ by defining
\[
X = E_{12} + E_{21} =
\frac{1}{\sqrt{3}}
  \left( \begin{matrix}
0 & 0 & 0 & 0 & 0& 0&0 &0 \\
0 & 1  & 0 & 0 & -1& 0&0 &0 \\
0 & 0 & -1 & 0 & 1& 0&0 &0 \\
0 & 0 & 0 & 0 & 0& 1 &-1 &0 \\
0 & -1 & 1 & 0 & 0& 0&0 &0 \\
0 & 0 & 0 & 1 & 0 & -1& 0 &0 \\
 0 & 0 & 0 & -1 & 0 & 0& 1  &0 \\
0 & 0 & 0 & 0 & 0 & 0&0 &0
\end{matrix}\right)
\]
\[
Y = -i E_{12} + i E_{21} =
\frac{i}{\sqrt{3}}
  \left( \begin{matrix}
0 & 0 & 0 & 0 & 0& 0&0 &0 \\
0 & 0  & -1 & 0 & 1& 0&0 &0 \\
0 & 1 & 0 & 0 & -1& 0&0 &0 \\
0 & 0 & 0 & 0 & 0& 1 &-1 &0 \\
0 & -1 & 1 & 0 & 0& 0&0 &0 \\
0 & 0 & 0 & -1 & 0 & 0& 1 &0 \\
 0 & 0 & 0 & 1 & 0 & -1& 0  &0 \\
0 & 0 & 0 & 0 & 0 & 0&0 &0
\end{matrix}\right)
\]
\[
Z = E_{11} - E_{22} =
\frac{1}{3}
  \left( \begin{matrix}
0 & 0 & 0 & 0 & 0& 0&0 &0 \\
0 & 1  & -2 & 0 & 1& 0&0 &0 \\
0 & -2 & 1 & 0 & 1& 0&0 &0 \\
0 & 0 & 0 & -2 & 0& 1 &1 &0 \\
0 & 1 & 1 & 0 & -2& 0&0 &0 \\
0 & 0 & 0 & 1 & 0 & 1& -2 &0 \\
 0 & 0 & 0 & 1 & 0 & -2& 1  &0 \\
0 & 0 & 0 & 0 & 0 & 0&0 &0
\end{matrix}\right) .
\]
One may verify directly that the anti-commutation relations are satisfied by $\{X,Y,Z\}$, and
that this triple belongs to $\fixed = \A^\prime =
\{\mathbf{X},\mathbf{Y},\mathbf{Z}\}^\prime$.

\section{The 4-Qubit Case}\label{S:4qubit}

In this section we determine the structure of $\A^\prime = \fixed$ for the collective noise
channel with $n=4$. We will show that $\A$ is unitarily equivalent to the unlinked orthogonal
direct sum
\[
\A \simeq \bbC \one_2 \oplus \big( \M_3 \otimes \one_3 \big)
\oplus \M_5,
\]
and hence
\[
\fixed = \A^\prime \simeq \M_2 \oplus \big( \M_3 \otimes \one_3
\big) \oplus \bbC \one_5.
\]
We can recognize a noiseless (decoherence-free) subspace
\cite{DG,LCW,ZR} which supports the first summand in the
decomposition of $\A$ and a noiseless subsystem in the second
summand. We establish the above forms of $\A$ and $\A^\prime$ by
identifying the maximal family of projections from Part~I of the
algorithm, $\{ P_{-3}, P_{-1}, P_{0,-1 / 3}, P_{0,1}, P_1, P_3
\}$, and show that: $P_{-3,} P_1$ are linked in $\A$ with $\rank
P_k = 1$
 for $k=-3,1$; the triple $P_{-1}, P_{0, -1 / 3}, P_3$ is linked in $\A$ with the rank of each
equal to three;  and $\rank P_{0,1} = 5$.

As for the 3-qubit case we may use elementary linear algebra to identify a set of vector space
generators for $\A^\prime$. It turns out that it is sufficient in this
4-qubit case to obtain a set of self-adjoint generators which have real
matrix entries. We find that $\A^\prime$ is
spanned by 10 such operators. For succinctness we shall only display the generators  used in
the spectral analysis. Let
\[
B_0 = B_0^\dagger =
  \begin{spmatrix}
0 & 0 & 0 & 0 & 0& 0&0 & 0 & 0 & 0 & 0 & 0 & 0& 0&0 &0 \\
0 & 0 & 0 & 0 & 0& 0&0 & 0 & 0 & 0 & 0 & 0 & 0& 0&0 &0 \\
0 & 0 & 0 & 0 & 1& 0&0 & 0 & -1 & 0 & 0 & 0 & 0& 0&0 &0 \\
0 & 0 & 0 & 0 & 0& 0&1 &0 & 0 & 0 & -1 & 0 & 0& 0&0 &0 \\
0 & 0 & 1 & 0 & 0& 0&0 & 0 & -1 & 0 & 0 & 0 & 0& 0&0 &0 \\
0 & 0 & 0 & 0 & 0& 0&1 &0 & 0 & 0 & 0 & 0 & -1& 0&0 &0 \\
0 & 0 & 0 & 1 & 0& 1&0 &0 & 0 & -2 & 0 & 0 & 0& 0&0 &0 \\
0 & 0 & 0 & 0 & 0& 0&0 &2 & 0 & 0 & 0 & -1 & 0& -1&0 &0 \\
0 & 0 & -1 & 0 & -1& 0&0 &0 & 2 & 0 & 0 & 0 & 0& 0&0 &0 \\
0 & 0 & 0 & 0 & 0& 0&-2 &0 & 0 & 0 & 1 & 0 & 1& 0&0 &0 \\
0 & 0 & 0 & -1 & 0& 0&0 &0 & 0 & 1 & 0 & 0 & 0& 0&0 &0 \\
0 & 0 & 0 & 0 & 0& 0&0 &-1 & 0 & 0 & 0 & 0 & 0& 1&0 &0 \\
0 & 0 & 0 & 0 & 0& -1&0 &0 & 0 & 1 & 0 & 0 & 0& 0&0 &0 \\
0 & 0 & 0 & 0 & 0& 0&0 &-1 & 0 & 0 & 0 & 1 & 0& 0&0 &0 \\
0 & 0 & 0 & 0 & 0& 0&0 &0 & 0 & 0 & 0 & 0 & 0& 0&0 &0 \\
0 & 0 & 0 & 0 & 0& 0&0 &0 & 0 & 0 & 0 & 0 & 0& 0&0 &0
\end{spmatrix}.
\]

This operator belongs to the commutant with characteristic polynomial
$\det (\lambda \one_{16} - B_0) = (\lambda + 3) (\lambda + 1)^3 \lambda^8 (\lambda - 1)
(\lambda - 3)^3$. Consider the set of spectral projections $\{P_{-3}, P_{-1}, P_0, P_1, P_3\}$
corresponding to the
eigenvalues $\lambda_1 = -3$, $\lambda_2 = -1$, $\lambda_3 = 0$, $\lambda_4 = 1$,
$\lambda_5 = 3$. These projections belong to $\A^\prime$ with $\rank P_{-3} = 1 = \rank P_1$,
$\rank P_{-1} = 3 = \rank P_3$, and $\rank P_0 = 8$. Let
\[
\left\{
\xi_{-3}  = \frac{1}{2\sqrt{3}}
\begin{sbmatrix}
0 & 0 & 0 & 1 & 0 & 1 & -2 & 0 & 0 & -2 & 1 & 0 & 1 & 0 & 0 & 0
\end{sbmatrix}^{t} \right.
\]
\[
\left\{
\xi_1 = \frac{1}{2}
\begin{sbmatrix}
0 & 0 & 0 & 1 & 0 & -1 & 0 & 0 & 0 & 0 & -1 & 0 & 1 & 0 & 0 & 0
\end{sbmatrix}^{t} \right.
\]
\[
\left\{
\begin{array}{rcl}
\xi_{-1,1} &=&  \frac{1}{\sqrt{2}}
\begin{sbmatrix}
0 & 0 & 0 & 0 & 0 & 0 & 0 & 0 & 0 & 0 & 0 & -1 & 0 & 1 & 0 & 0
\end{sbmatrix}^{t} \\
\xi_{-1,2} &=& \frac{1}{2}
\begin{sbmatrix}
0 & 0 & 0 & -1 & 0 & 1 & 0 & 0 & 0 & 0 & -1 & 0 & 1 & 0 & 0 & 0
\end{sbmatrix}^{t} \\
\xi_{-1,3} &=& \frac{1}{\sqrt{2}}
\begin{sbmatrix}
0 & 0 & -1 & 0 & 1 & 0 & 0 & 0 & 0 & 0 & 0 & 0 & 0 & 0 & 0 & 0
\end{sbmatrix}^{t}
\end{array} \right.
\]
\[
\left\{
\begin{array}{rcl}
\xi_{3,1} &=& \frac{1}{\sqrt{6}}
\begin{sbmatrix}
0 & 0 & -1 & 0 & -1 & 0 & 0 & 0 & 2 & 0 & 0 & 0 & 0 & 0 & 0 & 0
\end{sbmatrix}^{t} \\
\xi_{3,2} &=& \frac{1}{2 \sqrt{3}}
\begin{sbmatrix}
0 & 0 & 0 & -1 & 0 & -1 & -2 & 0 & 0 & 2 & 1 & 0 & 1 & 0 & 0 & 0
\end{sbmatrix}^{t} \\
\xi_{3,3} &=& \frac{1}{\sqrt{6}}
\begin{sbmatrix}
0 & 0 & 0 & 0 & 0 & 0 & 0 & -2 & 0 & 0 & 0 & 1 & 0 & 1 & 0 & 0
\end{sbmatrix}^{t}.
\end{array} \right.
\]
Then these projections may be computed as
\begin{align*}
P_{-3} &= \ket{\xi_1}\bra{\xi_1} \\
P_1 &= \ket{\xi_4}\bra{\xi_4} \\
P_{-1} &= \ket{\xi_{2,1}}\bra{\xi_{2,1}} +   \ket{\xi_{2,2}}\bra{\xi_{2,2}}
+ \ket{\xi_{2,3}}\bra{\xi_{2,3}} \\
P_3 &= \ket{\xi_{5,1}}\bra{\xi_{5,1}} +   \ket{\xi_{5,2}}\bra{\xi_{5,2}}
+ \ket{\xi_{5,3}}\bra{\xi_{5,3}} \\
P_0 &= \one_{16} - P_{-3} - P_{-1} - P_1 - P_3
\end{align*}

As rank one projections, $P_{-3} $ and $P_1$ are minimal in $\A^\prime$. Further,
compressing
the generators of $\A^\prime$ to $P_{-1}$ and $P_3$ reveals that $P\A^\prime P = \bbC P$ for
both $P=P_{-1}$ and $P= P_3$. Hence $P_{-1}$ and $P_3$ belong to the maximal family.  On
the other hand, the projection $P_0$ is not minimal $\A$-reducing. There are a number of ways
to see this, including verifying that $P_0\A^\prime P_0 \neq \bbC P_0$ directly, but perhaps the
easiest way is to observe that there are vector space generators for $\A^\prime$ which have all
their eigenvalue multiplicities strictly less than $8 = \rank P_0$. We
could find the (unique) maximal family of projections by
conducting a spectral analysis on any one of the non-scalar
vector space generators of $\A^\prime$.  Only two of the ten generators $B$ for $\A^\prime$
satisfy $P_0 B P_0 \neq 0$. They are:
\[
B_1 = B_1^\dagger =
  \begin{spmatrix}
1 & 0 & 0 & 0 & 0& 0&0 & 0 & 0 & 0 & 0 & 0 & 0& 0&0 &0 \\
0 & 0 & 0 & 0 & 0& 0&0 & 0 & 1 & 0 & 0 & 0 & 0& 0&0 &0 \\
0 & 0 & 0 & 0 & 0& 0&0 & 0 & 1 & 0 & 0 & 0 & 0& 0&0 &0 \\
0 & 0 & 0 & 0 & 0& 0&-1 &0 & 0 & 0 & 1 & 0 & 1& 0&0 &0 \\
0 & 0 & 0 & 0 & 0& 0&0 & 0 & 1 & 0 & 0 & 0 & 0& 0&0 &0 \\
0 & 0 & 0 & 0 & 0& 0&-1 &0 & 0 & 0 & 1 & 0 & 1& 0&0 &0 \\
0 & 0 & 0 & -1 & 0& -1&1 &0 & 0 & 2 & 0 & 0 & 0& 0&0 &0 \\
0 & 0 & 0 & 0 & 0& 0&0 &-2 & 0 & 0 & 0 & 1 & 0& 1&1 &0 \\
0 & 1 & 1 & 0 & 1& 0&0 &0 & -2 & 0 & 0 & 0 & 0& 0&0 &0 \\
0 & 0 & 0 & 0 & 0& 0&2 &0 & 0 & 1 & -1 & 0 & -1& 0&0 &0 \\
0 & 0 & 0 & 1 & 0& 1&0 &0 & 0 & -1 & 0 & 0 & 0& 0&0 &0 \\
0 & 0 & 0 & 0 & 0& 0&0 &1 & 0 & 0 & 0 & 0 & 0& 0&0 &0 \\
0 & 0 & 0 & 1 & 0& 1&0 &0 & 0 & -1 & 0 & 0 & 0& 0&0 &0 \\
0 & 0 & 0 & 0 & 0& 0&0 &1 & 0 & 0 & 0 & 0 & 0& 0&0 &0 \\
0 & 0 & 0 & 0 & 0& 0&0 &1 & 0 & 0 & 0 & 0 & 0& 0&0 &0 \\
0 & 0 & 0 & 0 & 0& 0&0 &0 & 0 & 0 & 0 & 0 & 0& 0&0 &1
\end{spmatrix}
\]
and
\[
B_2 = B_2^\dagger =
  \begin{spmatrix}
0 & 0 & 0 & 0 & 0& 0&0 & 0 & 0 & 0 & 0 & 0 & 0& 0&0 &0 \\
0 & 1 & 0 & 0 & 0& 0&0 & 0 & -1 & 0 & 0 & 0 & 0& 0&0 &0 \\
0 & 0 & 0 & 0 & 0& 0&0 & 0 & 0 & 0 & 0 & 0 & 0& 0&0 &0 \\
0 & 0 & 0 & 0 & 0& 0&1 &0 & 0 & 1 & -1 & 0 & -1& 0&0 &0 \\
0 & 0 & 0 & 0 & 0& 0&0 & 0 & 0 & 0 & 0 & 0 & 0& 0&0 &0 \\
0 & 0 & 0 & 0 & 0& 0&1 &0 & 0 & 1 & -1 & 0 & -1& 0&0 &0 \\
0 & 0 & 0 & 1 & 0& 1&-2 &0 & 0 & -2 & 1 & 0 & 1& 0&0 &0 \\
0 & 0 & 0 & 0 & 0& 0&0 &1 & 0 & 0 & 0 & 0 & 0& 0&-1 &0 \\
0 & -1 & 0 & 0 & 0& 0&0 &0 & 1 & 0 & 0 & 0 & 0& 0&0 &0 \\
0 & 0 & 0 & 1 & 0& 1&-2 &0 & 0 & -2 & 1 & 0 & 1& 0&0 &0 \\
0 & 0 & 0 & -1 & 0& -1&1 &0 & 0 & 1 & 0 & 0 & 0& 0&0 &0 \\
0 & 0 & 0 & 0 & 0& 0&0 &0 & 0 & 0 & 0 & 0 & 0& 0&0 &0 \\
0 & 0 & 0 & -1 & 0& -1&1 &0 & 0 & 1 & 0 & 0 & 0& 0&0 &0 \\
0 & 0 & 0 & 0 & 0& 0&0 &0 & 0 & 0 & 0 & 0 & 0& 0&0 &0 \\
0 & 0 & 0 & 0 & 0& 0&0 &-1 & 0 & 0 & 0 & 0 & 0& 0&1 &0 \\
0 & 0 & 0 & 0 & 0& 0&0 &0 & 0 & 0 & 0 & 0 & 0& 0&0 &0
\end{spmatrix}.
\]

Following the algorithm, we shall consider the spectral projections for $P_0 B_1 P_0$. We
compute $\det (\lambda \one_{16} - P_0 B_1 P_0) = (\lambda + 1 / 3)^3 \lambda^8 (\lambda -
1)^5$. Let $P_{0,-1 /3}, P_{0,0}, P_{0,1}$ be the projections onto the eigenspaces for
$\lambda_1
= - 1 / 3$,  $\lambda_2 = 0$,  $\lambda_3 = 1$. Since $P_0^\perp \leq P_{0,0}$ and $\rank
P_0^\perp = 8 = \rank P_{0,0}$, we have $P_{0,0} = P_0^\perp$, and we need only focus on
$P_{0,-1 / 3}$ and $P_{0,1}$. Observe that $\rank P_{0,-1 / 3} = 3$ and $\rank P_{0,1} = 5$.
Let
\[
\eta_1  = \frac{1}{2\sqrt{3}}
\begin{sbmatrix}
0 & 0 & 0 & 0 & 0 & 0 & 0 & 1 & 0 & 0 & 0 & 1 & 0 & 1 & -3 & 0
\end{sbmatrix}^{t}
\]
\[
\eta_2  = \frac{1}{\sqrt{6}}
\begin{sbmatrix}
0 & 0 & 0 & -1 & 0 & -1 & 1 & 0 & 0 & -1 & 1 & 0 & 1 & 0 & 0 & 0
\end{sbmatrix}^{t}
\]
\[
\eta_3  = \frac{1}{2\sqrt{3}}
\begin{sbmatrix}
0 & -3 & 1 & 0 & 1 & 0 & 0 & 0 & 1 & 0 & 0 & 0 & 0 & 0 & 0 & 0
\end{sbmatrix}^{t}
\]
Then $P_{0,-1 / 3}$ and $P_{0,1}$ may be computed as
\begin{align*}
P_{0,-1 / 3} &= \ket{\eta_1}\bra{\eta_1} + \ket{\eta_2}\bra{\eta_2} + \ket{\eta_3}\bra{\eta_3}
\\
P_{0,1} &= P_0 - P_{0,-1 / 3}.
\end{align*}

For the generators $B$ of $\A^\prime$ other than $B_1, B_2$ we have
\[
P_{0,-1 / 3} (B) P_{0,-1 / 3} =
P_{0,-1 / 3} (P_0 B P_0) P_{0,-1 / 3} = 0,
\]
and similarly $P_{0,1} B P_{0,1} = 0$. Moreover,
\[
P_{0,-1 / 3} (B_1) P_{0,-1 / 3} = - \frac{1}{3} P_{0,-1 / 3} \quad \quad
P_{0,-1 / 3} (B_2) P_{0,-1 / 3} =  \frac{4}{3} P_{0,-1 / 3}
\]
and
\[
P_{0,1} (B_1) P_{0,1} =  P_{0,1} \quad\quad
P_{0,1} (B_2) P_{0,1} = 0.
\]
Thus $P \A^\prime P = \bbC P$ for both $P= P_{0,-1 / 3}$ and $P= P_{0,1}$, and hence
$P_{0,-1 / 3}$
and $P_{0,1}$ are minimal $\A$-reducing projections.

As $ \one_{16} = P_{-3} + P_{-1} + P_{0,-1 / 3} + P_{0,1} + P_1 +P_3$, the
set of minimal $\A$-reducing projections
$\{P_{-3}, P_{-1} , P_{0,-1 / 3} , P_{0,1} , P_1 ,P_3\}$ does indeed form the maximal family
of projections we seek.   It
remains to find links. The commutators $[P_{-3}, B]$ for generators $B$ of $\A^\prime$ are not
all
zero. Hence, as the only rank one projections in this family, we may deduce that $P_{-3}$ and
$P_1$ are linked inside $\A$. Thus, $P_{-3} + P_1 \in (\A^\prime)^\prime = \A$ and
the block $\A(P_{-3} + P_1) = (P_{-3} + P_1) \A \simeq \bbC \one_2$ so that
\[
\A^\prime (P_{-3} + P_1) = (P_{-3} + P_1) \A^\prime \simeq \M_2.
\]
Further, a check of commutators $[P,B] $ for generators $B$ of
$\A^\prime$ with $P = P_{-1}, P_{3}, P_{0 -1 / 3}$
shows that none of these projections belongs to $(\A^\prime)^\prime = \A$.
It follows that
$P_{-1}, P_3, P_{0, -1 / 3}$ are linked inside $\A$, and thus
\begin{eqnarray*}
\A (P_{-1} + P_3 + P_{0, -1 / 3}) &\simeq& \M_3 \otimes \one_3 \\
&\simeq&  \one_3 \otimes \M_3 \\
&\simeq& \big( \M_3 \otimes \one_3
\big)^\prime \simeq \A^\prime (P_{-1} + P_3 + P_{0, -1 / 3}).
\end{eqnarray*}
Finally, we have $\A P_{0,1} \simeq \M_5$ and $\A^\prime P_{0,1} \simeq (\M_5)^\prime =
\bbC \one_5$, and the structure of $\fixed = \A^\prime$ is now apparent.

We finish by identifying copies of the Pauli matrices belonging to the matrix blocks inside
$\A^\prime = \fixed$.
For the block $\M_2\simeq \A^\prime (P_{-3} + P_1)$, we may define matrix units in
$\A^\prime$ by
\[
\begin{array}{ll}
E_{11} =P_{-3} = \ket{\xi_{-3}}\bra{\xi_{-3}} &
E_{12}  = \ket{\xi_{-3}}\bra{\xi_1} \\
E_{21} = \ket{\xi_1}\bra{\xi_{-3}} &
E_{22} = P_1= \ket{\xi_1}\bra{\xi_1}
\end{array}
\]
In particular, we may define the following operators  inside $\A^\prime (P_{-3} + P_1)
\subseteq \A^\prime = \fixed$:
\[
X = E_{12} + E_{21}, \quad Y = -i E_{12} + iE_{21}, \quad Z = E_{11} - E_{22}.
\]
We may also obtain matrix units  inside
$\A^\prime$ for the copy of $\M_3 \otimes \one_3
\simeq \A^\prime (P_{-1} + P_3 + P_{0, -1 / 3}) \subseteq \A^\prime$. For instance, a set of
$2\times 2$ matrix units inside this block is given by:
\[
\begin{array}{ll}
F_{11} =P_{-1} = \sum_{k=1}^3 \ket{\xi_{-1,k}}\bra{\xi_{-1,k}} &
F_{12} = \sum_{k=1}^3 \ket{\xi_{-1,k}}\bra{\xi_{3,k}} \\
F_{21} = \sum_{k=1}^3 \ket{\xi_{3,k}}\bra{\xi_{-1,k}} &
F_{22} =P_3 = \sum_{k=1}^3 \ket{\xi_{3,k}}\bra{\xi_{3,k}}
\end{array}
\]
Thus, a second copy of the Pauli matrices  inside $\A^\prime =
\fixed$ is realized by defining
\[
X = F_{12} + F_{21}, \quad Y = -i F_{12} + iF_{21}, \quad Z = F_{11} - F_{22}.
\]

\section{Conclusion}

Given a unital quantum channel $\Phi$, we have derived a
constructive proof for computing the explicit algebra structure,
as in (\ref{spatialform}), of the associated interaction algebra
$\A$ and noise commutant $\A^\prime$.  We mention that Zarikian
\cite{Zar} has recently written Matlab algorithms which also
utilizes much of the theory discussed here to compute, among
other things, the structures of such operator algebras. An
important subtlety in this process involves the detection of
ampliations within these algebras which arise from linked minimal
projections. In the case of non-unital noise the algorithm may be
easily adapted to find the structures of $\A$ and $\A^\prime$,
since both are $\dagger$-algebras by definition, but the
connection with the fixed point set $\fixed = \A^\prime$ is no
longer valid.

As illustrations we worked through the process for several simple
channels and the $n=3$ and $n=4$ qubit cases of the collective
noise channels which arise from collective rotations. We have also
used our method to compute the noise commutant for the 5-qubit
case, it is given by $\A^\prime \simeq \bbC \one_6 \oplus
(\M_4\otimes \one_4) \oplus (\M_5 \otimes \one_2)$.

While the algorithm is capable of analyzing any situation, in
particular cases a more specialized and conceptual investigation
may be necessary. For instance, when $n>>0$ the computations in
the algorithm required to compute the noise commutant of the
$n$-qubit collective noise channel make the problem infeasible. In
forthcoming work, the authors and Poulin \cite{HKLP} use such a
conceptual investigation to derive the explicit structure of the
noise commutant for the general $n$-qubit case of the collective
noise channels arising from collective rotation.









{\noindent}{\it Acknowledgements.} We thank the referee for
several helpful comments.  We are grateful to the Perimeter
Institute for providing resources which helped foster this
collaboration. We also acknowledge funding from NSERC, MITACS, and
ARDA. The second author would like to thank Michele Mosca of the
Institute for Quantum Computing and members of  the Department of
Mathematics at Purdue University for kind hospitality during
recent visits.


%

\end{document}